\documentclass[10pt,journal,compsoc]{IEEEtran}

\usepackage{booktabs} 
\usepackage{xcolor,colortbl}

\usepackage{resizegather}
\usepackage[export]{adjustbox}
\usepackage{amsmath}
\usepackage{dblfloatfix}
\usepackage{tabularx}
\usepackage{multirow}
\usepackage{enumitem}
\usepackage{standalone}
\usepackage{algorithm}
\usepackage{amsfonts}
\usepackage{makecell}

\usepackage[listings,skins,breakable]{tcolorbox}
\usepackage{algpseudocode}
\usepackage{tikz,pgfplots}
\usepackage{subcaption}
\usepackage{pgfplots}
\usepackage{makecell}
\usepackage{url}
\usepackage{hhline}
\usepgfplotslibrary{fillbetween}
\usepgfplotslibrary{statistics}
\usetikzlibrary{backgrounds}
\usetikzlibrary{pgfplots.groupplots}
\usepackage{pgf}
\usepgflibrary{plothandlers,plotmarks}
\usepackage{csquotes}
\usepackage{float}
\usepackage{pgf-pie}
\usepackage[flushleft]{threeparttable}
\pgfplotsset{compat=newest}
\pgfplotsset{
        /pgfplots/ybar legend/.style={
        /pgfplots/legend image code/.code={%
        \draw[##1,/tikz/.cd,bar width=3pt,yshift=-0.2em,bar shift=0pt]
                plot coordinates {(0cm,0.8em)};},
},
}

\newbox\aMark
\setbox\aMark\hbox{\begin{pgfpicture}\textcolor{blue}{\pgfuseplotmark{o}}\end{pgfpicture}}

\newbox\bMark
\setbox\bMark\hbox{\begin{pgfpicture}\textcolor{red}{\pgfuseplotmark{star}}\end{pgfpicture}}

\newcolumntype{P}[1]{>{\centering\arraybackslash}m{#1}}
\newcolumntype{Y}{>{\centering\arraybackslash}X}

\newlength{\bibitemsep}
\newlength{\bibparskip}
\let\oldthebibliography\thebibliography
\renewcommand\thebibliography[1]{%
  \oldthebibliography{#1}%
  \setlength{\parskip}{\bibitemsep}%
  \setlength{\itemsep}{\bibparskip}%
}

\newcommand{\rev}[1]{\textcolor{black}{#1}}
\ifCLASSOPTIONcompsoc
  \usepackage[nocompress]{cite}
\else
  \usepackage{cite}
\fi

\ifCLASSINFOpdf
\else
\fi

\begin{document}
%
\title{Search-Based Software Engineering for Self-Adaptive Systems: Survey, Disappointments, Suggestions and Opportunities}

\author{Tao~Chen,
        Miqing~Li,
        Ke~Li,
        and~Kalyanmoy~Deb,~\IEEEmembership{Fellow,~IEEE}
\IEEEcompsocitemizethanks{\IEEEcompsocthanksitem Tao Chen is with the Department
of Computer Science, Lougborough University,
UK. E-mail: t.t.chen@lboro.ac.uk\protect
\IEEEcompsocthanksitem Miqing Li is with the School
of Computer Science, University of Birmingham,
UK. E-mail: m.li.8@bham.ac.uk\protect
\IEEEcompsocthanksitem Ke Li is with the Department
of Computer Science, University of Exeter,
UK. E-mail: k.li@exeter.ac.uk\protect
\IEEEcompsocthanksitem Kalyanmoy Deb is with the Department of Computer and Engineering, Michigan State University, United States. E-mail: kdeb@msu.edu\protect\\

}
\thanks{Manuscript received April 19, 2005; revised August 26, 2015.}}

%

\IEEEtitleabstractindextext{%
\begin{abstract}
Search-Based Software Engineering (SBSE) is a promising paradigm that exploits the computational search to optimize different processes when engineering complex software systems. Self-adaptive system (SAS) is one category of such complex systems that permits to optimize different functional and non-functional objectives/criteria under changing environments (e.g., requirements and workload), which involves problems that are subject to search. In this regard, over years, there has been a considerable amount of work that investigates SBSE for SASs. In this paper, we provide the first systematic and comprehensive survey exclusively on SBSE for SASs, covering papers in 27 venues from 7 repositories, which eventually leads to several key statistics from the most notable 74 primary studies in this particular field of research. Our results, surprisingly, have revealed five disappointments that are of utmost importance and can result in serve consequences but have been overwhelmingly ignored in existing studies. We provide theoretical and/or experimental evidence to justify our arguments against the disappointments, present suggestions, and highlight the promising research opportunities towards their mitigation. We also elaborate on three other emergent, but currently under-explored opportunities for future work on SBSE for SASs. By mitigating the disappointments revealed in this work, together with the highlighted opportunities, we hope to be able to excite a much more significant growth in this particular research direction.
\end{abstract}

}

\maketitle

\IEEEdisplaynontitleabstractindextext

%
\IEEEpeerreviewmaketitle

\IEEEraisesectionheading{\section{Introduction}\label{sec:intro}}


\IEEEPARstart{E}{ngineering} software systems with the ability to reason and adapt themselves under changes (e.g., on its states, requirements, and the environment) has emerged as a successful paradigm for handling runtime dynamics and uncertainty. The resulted software system, namely self-adaptive systems (SASs), has become one of the most complex artifacts that have ever been created by humans. Many complex software systems require optimization in the engineering process and the SASs are of no exception. For example, the configuration of SASs' adaptable parameters is a very typical optimization problem in which the best-configured values (and possibly sequence) need to be searched in order to achieve optimality on different functional and non-functional objectives/criteria~\cite{DBLP:journals/taas/SalehieT09}. However, optimizing SASs is important yet challenging, as human intervention is profoundly limited and there may be an explosion of the possible adaptation solutions, together with multiple conflicting objectives under resource constraints and feature dependencies~\cite{DBLP:conf/dagstuhl/LemosGMSA}. As a result, intelligent search is required to fulfill the requirement of optimization in various domains of SASs. 

Search-Based Software Engineering (SBSE)~\cite{DBLP:journals/csur/HarmanMZ12} is one example of a form of ‘Intelligent Software Engineering’ that has been widely applied across many software engineering domains that demand optimization, including requirements~\cite{DBLP:journals/infsof/ZhangHL13}, design~\cite{DBLP:journals/tse/PraditwongHY11}, testing~\cite{DBLP:conf/icst/FraserA12} and refactoring~\cite{DBLP:journals/tse/LuWYAN19}. Specifically, SBSE applies computational search, i.e., various search algorithms, to automatically and dynamically seek solutions for minimizing/maximizing objective(s) or for satisfying certain constraint(s) in software engineering. In particular, SBSE can be either single-objective, where a single fitness would be used to guide the search that leads to a single optimal solution; or multiple objectives, in which the search is steered by either a weighted aggregation (a.k.a. utility-driven search) or a pressure to approximate the Pareto front~\cite{Ehrgott2006}, i.e., Pareto search\footnote{Pareto search refers to any algorithms that seek to search for the Pareto front in the presence of multiple objectives, including Pareto-dominance based, indicator-based and decomposition-based multi-objective algorithms.}~\cite{collette2013multiobjective}.

Over years, there have been some successful attempts on exploring SBSE for SASs~\cite{DBLP:conf/icac/RamirezKCM09,DBLP:journals/jss/PascualLPFE15,DBLP:conf/icse/KinneerCWGG18,DBLP:journals/infsof/ChenLY19,DBLP:journals/tosem/ChenLBY18}. Indeed, as pointed out by Harman et al.~\cite{DBLP:conf/icse/HarmanJLPMYW14}, the very natural requirement of dynamic and automated reasoning in SAS provides a perfect place for SBSE, which targets exactly such a need. Nevertheless, the work in such direction is still arguably much less active compared with the other problems of software engineering, e.g., software testing~\cite{DBLP:journals/csur/HarmanMZ12}, where SBSE has become a standard. We believe that one of the reasons for this is because, to the best of our knowledge, there has been no explicit survey on the topic of SBSE for SASs. As such, we lack a general overview, and hence both the SBSE and SAS practitioners are struggling to understand, e.g., what search algorithms to use and how they can be tailored, in what contexts, and how the search results can be assessed. This is what originally motivates this paper, in which we aim to bridge such gap by conducting a systematic survey on papers over 27 venues and 7 repositories, based on which 409 ones were identified for detailed review and eventually 74 primary studies were extracted for the analysis.

The survey has in fact led to a surprising result: we have identified five disappointing phenomena in the current research on SBSE for SASs, some of which can pose immediate threats to the validity in current research while others can negatively affect the field in the long-term. We would like to stress that we term them ``disappointments" because they, albeit very likely, may not always lead to an issue for an individual study. However, from the perspective of the entire field, these disappointments bear a resemblance to the ``bad smells" in software engineering analogy --- phenomena where there are hints that suggest there can be an issue. For example, randomly choosing a search algorithm without justifying with the specifics of the SAS problem may work when the choice happens to be the suitable one, this is nevertheless not ideal if it becomes an overwhelming phenomenon for the field. The presence of those disappointments is perhaps another reason that prevents a significant growth of the research on SBSE for SASs. To further advance this direction of research and mitigate the disappointments discovered, we provide suggestions and highlight eight promising research opportunities that are currently under-explored in the existing work from the literature. 

To the best of our knowledge, our work is the very first endeavor that aims to target explicitly on SBSE for SASs, offering a comprehensive overview and critical analysis of this field. Specifically, our contributions in this paper are three folds:

\begin{enumerate}
\item We conduct a systematic survey of the work on SBSE for SASs published between 2009 and 2019. The research questions (RQs) that our survey aims to answer are:

\begin{itemize}
\item[---]\textbf{RQ1:} What, where, and why search algorithms are used for SAS?
\item[---]\textbf{RQ2:} What, how, and why SAS objectives are defined and handled?
\item[---]\textbf{RQ3:} What and why evaluation methods are used to assess the results on Pareto search for SASs?
\item[---]\textbf{RQ4:} What, how, and why domain information is used to specialize the search algorithm for SAS?
\item[---]\textbf{RQ5:} What, why, and how many subject SASs are studied in order to generalize the conclusion drawn?
\end{itemize}

\item Drawing on the survey results for the above RQs, we have identified five disappointing phenomena in current work, for which we discuss the disappointments supported with theoretical and/or experimental justification of the possible issues.

\item We provide suggestions and highlight eight promising research opportunities in SBSE for SASs, some of which are promising to mitigate the disappointments identified and discuss their challenges.
\end{enumerate}

Noteworthily, our goal is not to question the importance and significance of existing work, but rather to summarize the key statistics that enable us to discuss, justify and raise debates about some of the overwhelmingly possible issues in existing studies related to SBSE for SASs, which have sadly disappointed us. We feel that respectful scientific debates are very important for sustainable research, particularly in such an interdisciplinary topic where research from the well-established community of SBSE and computational optimization may still be relatively new to the SAS practitioners. Indeed, explicit debates on topics may timely reveal the opposing ideas and can often excite significant growth of the research field (e.g., see~\cite{DBLP:conf/icse/Li0Y18}). By addressing those disappointments, together with promising research opportunities that are currently under-explored in SBSE for SASs, we envisage to further grow this particular research field.

The remaining of this paper is organized as follows. Section~\ref{sec:bg} introduces some background information of SBSE and SASs. Section~\ref{sec:method} presents the research methodology following by detailed elaboration on our literature review protocol in Section~\ref{sec:review}. Section~\ref{sec:rq} analyzes the results obtained from our systematic survey with respect to our RQs, discusses the disappointments with justification, and highlights the suggestions and research opportunities for mitigation. Other currently under-explored opportunities in SBSE for SASs are discussed in Section~\ref{sec:opp}. Threats to validity and conclusions are included in Section~\ref{sec:tov} and~\ref{sec:con}, respectively. 

\section{Preliminaries}
\label{sec:bg}

\subsection{Search-Based Software Engineering}




In SBSE, the behaviors of search are specifically referring to specialize a metaheuristic search algorithm (evolutionary algorithm in particular) within a search space of candidate solutions, guided by a fitness function that evaluates the quality of solutions, with an aim to find the optimal or near-optimal one(s)~\cite{DBLP:journals/csur/HarmanMZ12}. According to the domain information, the fundamental tasks of specializing SBSE to a SE problem (in fact, any optimization problem), as discussed by Harman et al.~\cite{DBLP:journals/csur/HarmanMZ12}, include reformulating the solution representation of the problem and designing the objective/fitness function(s) that distinguishes between good and bad solutions. 


The successful application of search algorithms for software engineering over years has led to an increasing interest in other forms of optimization for software engineering that are not necessarily directly based on a metaheuristic search. Indeed, from the literature, it is not uncommon to find SBSE, or simply search algorithms, applied to any form of optimization in which the problem comes from software engineering and the solutions are subject to search. In this paper, we, therefore, include classical Operational
Research and Computational Optimzation techniques, as well as the metaheuristic search algorithms in the traditional understanding of SBSE.


One 
important distinction on the problems of SBSE is whether the software engineering problem involves multiple conflicting objectives. 
In the single-objective case, 
the search can be directly guided and evaluated by a fitness function. 
In contrast, 
when multiple objectives are involved, 
the search and evaluation become much more complex due to the presence of conflicts, 
i.e., trade-offs are required. 
Indeed, 
any multi-objective problem may be converted into a single-objective one via a certain form of weighted aggregation. 
However, 
as we will show in Section~\ref{sec:rq2}, 
this does not come without any cost: 
the precise quantification of weights can be very difficult, 
if not impossible, 
and the conflicting relation between objectives may be blurred, 
causing some ideal solutions hard to be found~\cite{DBLP:journals/csur/HarmanMZ12}. 
This is the fact that has been well accepted by the SBSE community in many different problems~\cite{DBLP:journals/tse/PraditwongHY11,zhang2007multi,DBLP:conf/issta/YooH07}.

In SBSE, a standard way of handling multi-objectivity, 
borrowed from the Computational Optimization and Evolutionary Computation community, 
is to use the notion of Pareto dominance and optimality~\cite{Ehrgott2006} 
by which the searched result is often a set of trade-off solutions instead of a single one. 
By definition, 
a solution \texttt{A} is said to be (Pareto) dominated by \texttt{B} 
if all of \texttt{B}'s objective values are better than or equals to the corresponding objective values of \texttt{A}, 
and there is a least one objective on which \texttt{B} is better than \texttt{A}. 
A solution is called Pareto optimal if it is not dominated by any solution in the search space. 
The set of all the Pareto optimal solutions is called the Pareto set 
while their images in the objective space constitute the Pareto front of the problem. 

Without additional information from the decision maker, 
the solutions in a nondominanted set are in fact incomparable. 
This has led to the distinctiveness of Pareto search algorithms, 
in which the search does not aim for a single (weighted) optimal solution,
but rather for a set of solutions that can well-represent the whole Pareto front. 
Such ``representation'' can be broken down into four aspects with respect to solution sets' quality:
convergence, spread, uniformity, and cardinality~\cite{DBLP:journals/csur/LiY19}
which various indicators are designed to evaluate. 
Convergence refers to the closeness of the solution set to the Pareto front;
spread considers the region of the set covering;
uniformity refers to the evenness of solutions distributed in the set;
and cardinality refers to the number of solutions in the set.   
A more thorough overview of the Pareto search algorithms in SBSE can be found in~\cite{DBLP:journals/csur/HarmanMZ12,DBLP:conf/splc/HarmanJKLPZ14,DBLP:journals/tse/RamirezRS19}.

	\begin{figure}[t!]
		\centering
		\includegraphics[width=0.8\columnwidth]{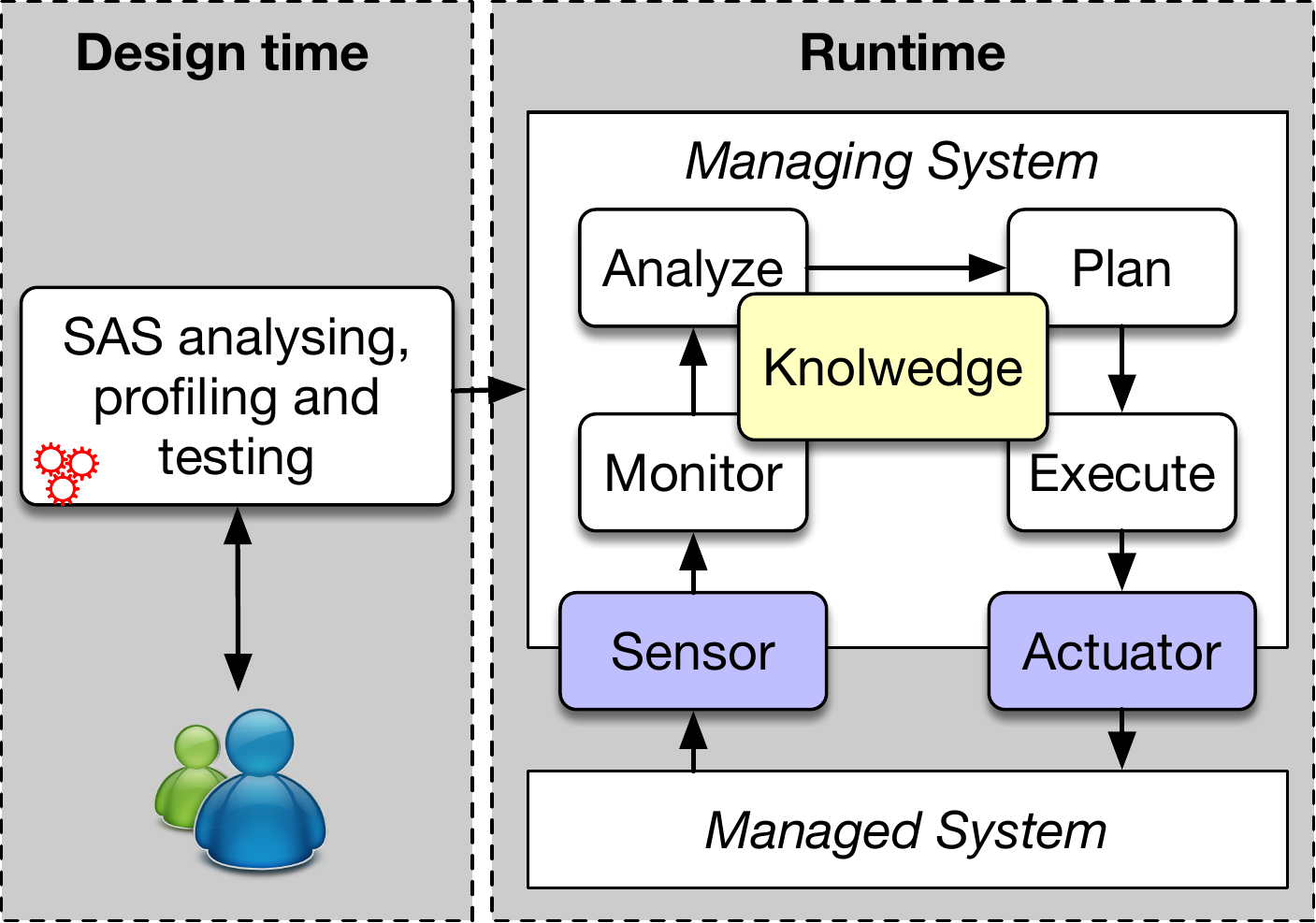}
		\caption{General overview of SAS.}
		\label{fig:sas}
	\end{figure}

\subsection{Self-Adaptive Systems}


The ever increasing complexity of engineering software systems has led to the high demand for software systems to be versatile, resilient, and dependable to changes in its operational contexts, requirements, and environments. This cannot be made possible without the notion of self-adaptation\textemdash the ability of a software system that permits it to modifies its own behaviors according to the perceptions about its interior states and the exterior factors, leading to a special type of software systems termed SASs.

Figure~\ref{fig:sas} shows a general overview of the SAS, deriving from the most widely-adopted MAPE-K architectural model~\cite{DBLP:journals/computer/KephartC03}. As can be seen, at runtime, there is a feedback loop consists of two key components~\cite{DBLP:conf/icse/WeynsIMA12}: a managed system that is adaptable, but does not itself have the ability to adapt; and a managing system that encapsulates all the core logic to realize self-adaptation for the managed system --- this is also the key component in which most work on SBSE for SASs lies, as we will show in Section~\ref{sec:rq1}. In particular, under MAPE-K, runtime self-adaptation in SAS is governed by the managing system via the following phases:


\begin{enumerate}

    \item \textbf{Monitor:} Collecting data on the managed system and the environment through sensors.
    \item \textbf{Analyze:} Making decisions about whether adaptation is required.
    \item \textbf{Plan:} Reasoning about the suitable adaptation solution.
    \item \textbf{Execute:} Physically conducting adaptations via actuators.
    \item \textbf{Knowledge:} Centralized abstraction of relevant aspects of the managed system, its environment, and the self-adaptation objectives. This is the shared phase by the other four.
    
\end{enumerate}

Although at the first glance of the term `self-adaptation', one may think that it is only related to the running software systems, the engineering processes of SASs are, in fact, spread over both design-time and runtime~\cite{DBLP:journals/taas/SalehieT09,DBLP:conf/dagstuhl/LemosGMSA}. As shown in Figure~\ref{fig:sas}, the design-time tasks for SAS can provide important insights to the process embedded in each MAPE-K phase~\cite{DBLP:conf/dagstuhl/LemosGMSA,DBLP:conf/kbse/GerasimouTC15,DBLP:journals/tosem/ChenLBY18}. Indeed, design-time profiling of the possible adaptation solution and their effect on the quality, under different environmental conditions, have been shown to be helpful for building more effective policies for runtime self-adaptation~\cite{DBLP:journals/tosem/ChenLBY18,DBLP:journals/ase/GerasimouCT18}. As a concrete example, Gerasimou et al.~\cite{ DBLP:journals/ase/GerasimouCT18} shows that design-time analysis can help to determine the strategy design in the \textit{Analysis} and \textit{Plan} phases. In this paper, we, therefore, include not only the work on SASs runtime but also the design-time studies of SASs as long as they serve as significant understandings for runtime self-adaptation. We would like to stress that although the design-time problem of SASs is aligned with our purpose in this work, we, however, do not consider those studies at design-time without referring to their implication on runtime self-adaptation. This is because the ultimate goal of SAS is to allow the software system to run and dynamically adapt according to the time-varying changes as they emerge.

Given the generic notions from Figure~\ref{fig:sas}, any adaptable/managed software systems can form a SAS, providing that some, if not all, of the internal parts (i.e., variation points) can be changed as the software system runs. In real-world scenarios, many widely used software systems are readily prepared for self-adaptation. For example, \texttt{MySQL}, which is one of the most popular Relational Database Management Systems, has around one-third of its variation points that can be changed at runtime\footnote{https://dev.mysql.com/doc/refman/5.5/en/server-system-variable-reference.html}. Because of this, from the software engineering research community, it is not uncommon to find that research of SASs has been conducted under different themes, one of the most noticeable examples is Dynamic Software Product Line~\cite{DBLP:journals/computer/BencomoHA12,baresi2014self,classen2008modelling}. In addition to software engineering, SAS research has been spread over the other communities, e.g., System Engineering, Service Computing, and Cloud Computing. Our survey, therefore, does not restrict only to software engineering research, but also to any other communities following our review protocol introduced in Section~\ref{sec:method}. More detailed surveys of the SASs in general are available from the literature, see~\cite{DBLP:journals/taas/SalehieT09} and~\cite{DBLP:conf/dagstuhl/LemosGMSA}.


\subsection{Marrying SBSE with SASs}

Self-adaptations of a SAS are certainly not being conducted for no reason, they are designed to serve a certain purpose: to improve the quality of software systems, including both functional and non-functional quality. While there are certain quality dimensions that are widely applicable to many domains, e.g., latency, throughput, and availability, the actual variation points by which a SAS can modify itself are highly diverse case by case.

Because of the above reasons, engineering SASs can be abstracted as developing automated and dynamic methods that involve tuning (searching) different parts or processes of the software systems, with an aim to improve functional or non-functional quality. This fits precisely with the purpose of SBSE and therefore raises a perfect marriage between the two fields. 


\section{Research Methodology}
\label{sec:method}

	\begin{figure}[t!]
		\centering
		\includegraphics[width=\columnwidth]{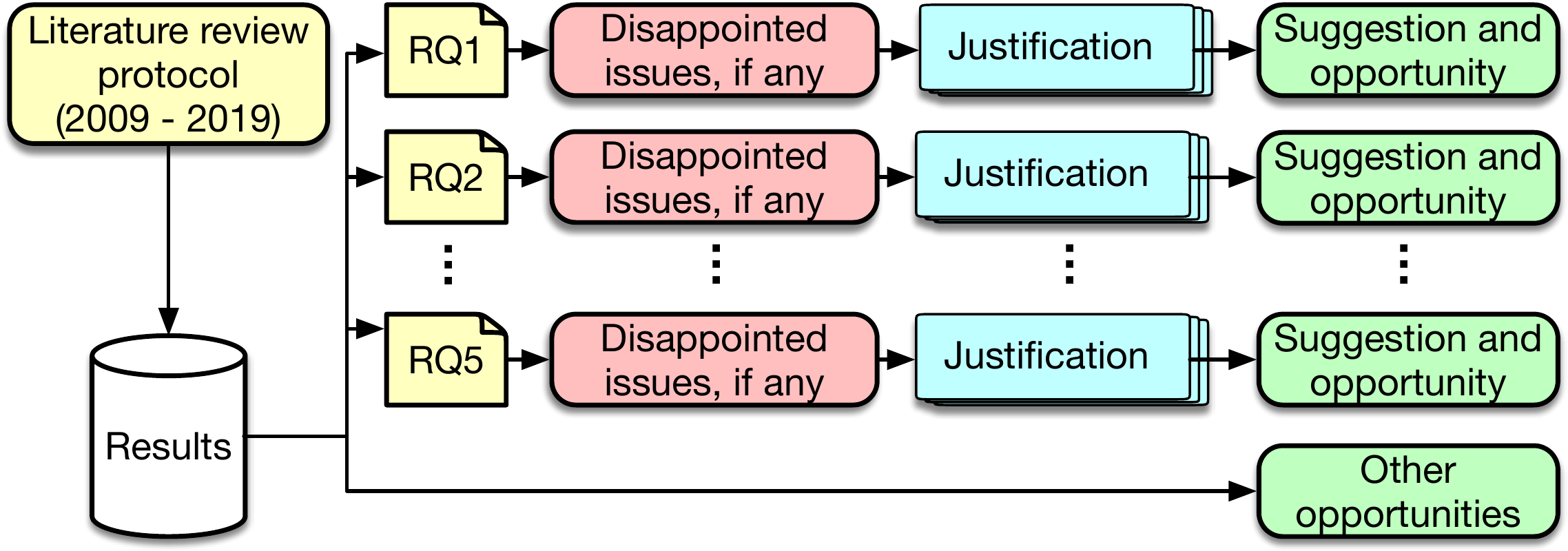}
		\caption{Overall research methodology.}
		\label{fig:research-method}
	\end{figure}

The overview of our research methodology in this work has been shown in Figure~\ref{fig:research-method}. As can be seen, to understand the state-of-the-art on exploring SBSE when engineering SASs, we first conducted a systematic literature review covering the papers published from 2009 to 2019. The reason why we chose 2009 as the starting year of the review is that it is the last year covered by two well-known surveys for the SBSE~\cite{DBLP:journals/csur/HarmanMZ12} and SAS domain~\cite{DBLP:journals/taas/SalehieT09}, respectively. Therefore, this work seeks to overcome such a gap and, also for the first time, uniquely focuses on how SBSE has been involved in the research on SAS over the last decade.

The review methodology follows the best practice of systematic literature review for software engineering~\cite{DBLP:journals/infsof/KitchenhamBBTBL09}, consisting of clear search protocol, inclusive/exclusive criteria, pragmatic classification of data items and formal data collection process. In brief, the review has two goals: (i) to provide summative statistics with respect to the aforementioned RQs; (ii) to identify the sources that derive our discussion on the disappointments and opportunities for this particular research field.

For each RQ, we discuss the results, identify disappointments (if any) together with their theoretical and/or experimental justification, provide suggestions and outline the future research opportunities that could potentially mitigate those disappointments. In addition to these, we discuss other opportunities that are currently under-explored in SBSE for SASs in general.

\section{Review Protocol Overview}
\label{sec:review}

\begin{figure}[t!]
		\centering
		\includegraphics[width=\columnwidth]{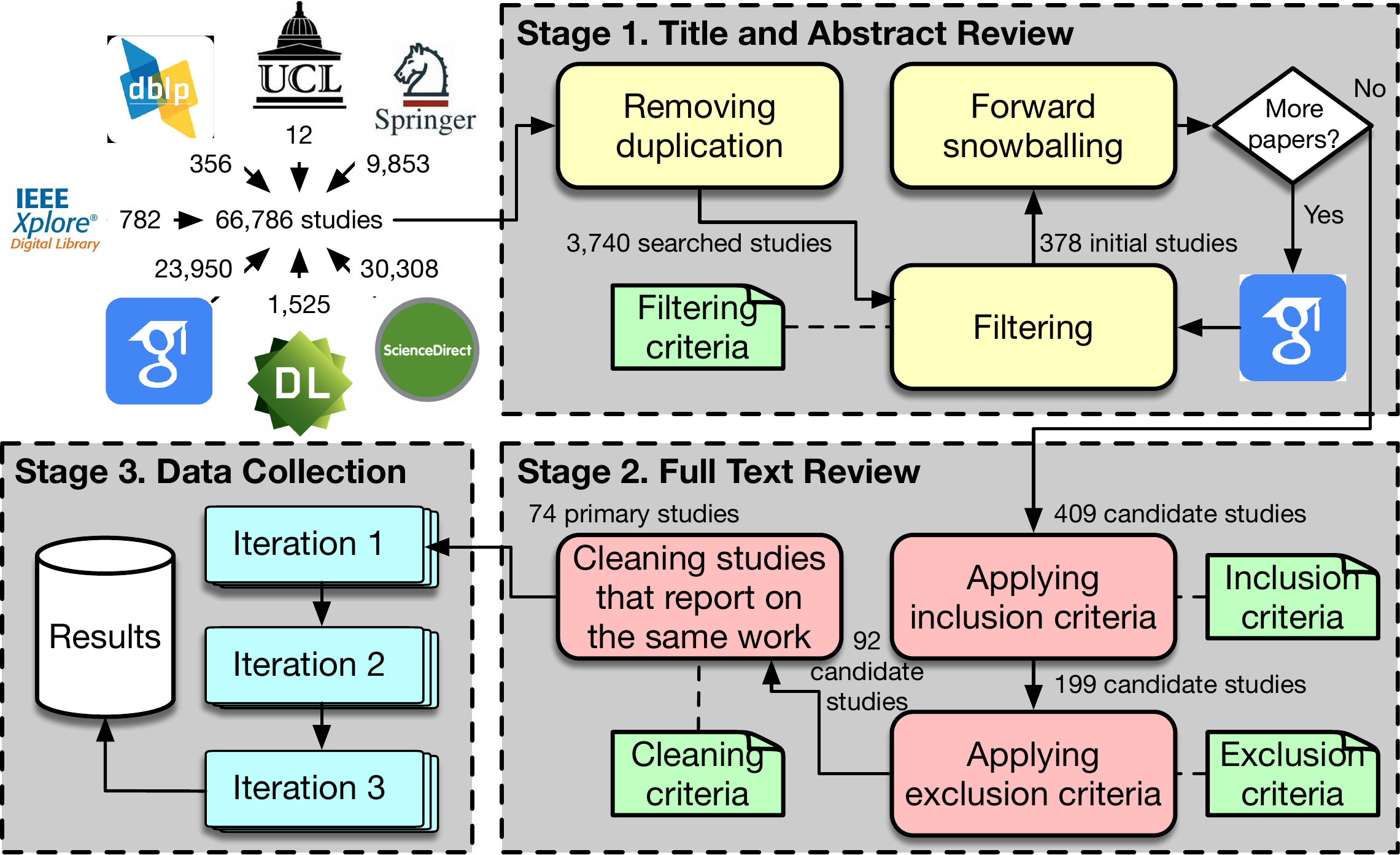}
		\caption{Systematic literature review protocol.}
		\label{fig:slr}
	\end{figure}

\rev{As shown in Figure~\ref{fig:slr}, our literature review protocol exploits automatic search to obtain a set of 66,786 studies from various sources (see Section~\ref{sec:scope}). Starting from stage 1, we removed duplication by automatically matching their titles\footnote{Patents, citation entries, inaccessible papers, and any other non-English documents were also eliminated.}, leading to 3,740 \textbf{\emph{searched studies}}. Next, we filtered the searched studies according to their titles and abstracts. A study was ruled out if it meets any of the two filtering criteria below:}
	\begin{itemize}
	    \item \rev{The paper is not relevant to SAS.}
	    \item \rev{The paper does not conduct research in the context of software or system engineering.}
	\end{itemize}
	
\rev{The filtering process resulted in a much smaller and more concise set of 378 \textbf{\emph{candidate studies}}. We then conducted a manual search using the iterative forward snowballing as suggested by Felizardo et al.~\cite{DBLP:conf/esem/FelizardoMKSV16}, where the newly included studies (after filtering) were placed into the next snowballing round. Note that we did not do backward snowballing because the studies searched in our work are restricted within the last decade, therefore the backward snowballing would too easily violate such a requirement of timeliness. To avoid a complicated process in the snowballing, we relied on Google Scholar as the single source therein following the best practice for software engineering surveys~\cite{DBLP:journals/tse/GalsterWTMA14}. The snowballing process stopped when no new studies can be found, leading to 409 candidate studies, and the procedure for full-text review begins thereafter.}
	
\rev{At stage 2, we reviewed all the 409 studies and temporarily keep some of them using the inclusion criteria from Section~\ref{sec:in-ex}, which resulted in 199 candidate studies. We then applied the exclusion criteria (see Section~\ref{sec:in-ex}) to extract the temporarily included studies, leading to 92 candidate studies. By using the cleaning criteria specified in Section~\ref{sec:in-ex}, a further cleaning process was conducted to prune different studies that essentially report on the same work, e.g., journal papers extended from a conference version. All the processes finally produced 74 \textbf{\emph{primary studies}} ready for data analysis and collection.}
	
\rev{For all primary studies at stage 3, we conducted a systematic and pragmatic data collection process that consists of three iterations, for which we elaborate in Sections~\ref{sec:item-class} and~\ref{sec:data-collection}.}

\subsection{Search String}
\label{sec:scope}

\rev{From 16th to 30th Sep 2019, we conducted an automatic search over a wide range of scientific literature sources, including ACM Library, IEEE Xplore, ScienceDirect, SpringerLink, Google Scholar, DBLP and the SBSE repository maintained by the CREST research group at UCL\footnote{http://crestweb.cs.ucl.ac.uk/resources/sbse\_repository}.}

The used search string was designed to cover a variety of computational search applied in the context of SASs. Synonyms and keywords were properly linked via logical connectors (AND, OR) to build the search term. The final search string is shown as below:

\begin{displayquote} 
\textit{(``optimization" OR ``search algorithm" OR ``search based"  OR ``multi-objective") AND (``adaptive software" OR ``adaptive system" OR ``dynamic software product line" OR ``autonomic")}
\end{displayquote}

The first set of terms before the AND connector consists of the common keywords for SBSE surveys~\cite{DBLP:journals/csur/HarmanMZ12,Sayyad2013b}, while the second contains terms that are commonly appeared in the SAS related surveys~\cite{DBLP:conf/c3s2e/WeynsIIA12,DBLP:conf/refsq/YangLJC14}. Noteworthily, we explicitly placed \textit{``dynamic software product line"} in the string because as far as we aware, this is the only domain that has been formally acknowledged of being highly relevant to both the SBSE~\cite{DBLP:conf/splc/HarmanJKLPZ14} and SAS~\cite{DBLP:journals/computer/BencomoHA12,baresi2014self,classen2008modelling} community. In this way, we retain a high degree of coverage as evidenced by the number of returned results from Figure~\ref{fig:slr}.

\rev{Using the above string, we conducted a full-text search on ACM Library, IEEE Xplore, ScienceDirect, SpringerLink, and Google Scholar, but rely on searching the title only for DBLP and UCL's SBSE repository, due to their restricted feature. Since DBLP does not work on the whole search string, we paired each term in the first bracket with each one from the second bracket. The results of all pairs were collected. Due to the similar reason, for the UCL's SBSE repository, we searched each term from the second bracket independently and collected all results returned, as it is known that all the studies in this source are SBSE related.}

\subsection{Inclusion, Exclusion, and Cleaning Criteria}
\label{sec:in-ex}
For the selected candidate studies, we first identify the primary ones by using the inclusion criteria as below; studies meeting all of the criteria were temporarily chosen:

\begin{enumerate}

\item The study specifies the deign, or application, of the computational search algorithm as a major part of the solution to a problem of engineering SASs. If this is not the case, the paper should at least present a formulation of the SAS problem that can be subject to computational search.

\item The study investigates problems related to SAS runtime, or it is a design-time problem that can provide significant insights for runtime self-adaptation of SASs with discussion.


\item The study explicitly or implicitly discusses, or at least made assumptions about, the generality of the problem and solution when engineering SASs to the wider context, despite that it may focus on a particular domain of SAS (e.g., Cloud, Services, Internet-of-Things and Cyber-Physical Systems). 


\item The problem in the study to be solved by SBSE is derived from a software or system engineering perspective.


\item The study includes quantitative experimental results with clear instructions on how the results were obtained.
		
\item The study uses at least one method or quality indicator to evaluate the experimental results. 

\end{enumerate}

Subsequently, studies meeting any of the exclusion criteria below are ruled out:

\begin{enumerate}

\item The study neither explicitly nor implicitly mentions SBSE, where the computational search is the key; or the search problem is not considered as an important part of the approach.


\item The study is not ``highly visible'' or widely followed. We used the citation information from Google Scholar as a single metric to (partially) assess the impact of a study\footnote{Admittedly, no single metric can well quantify the impact of a paper. Nevertheless, 
the citation count can tell something about a paper, e.g., its popularity.}. In particular, we follow a pragmatic strategy that: a study has 5 citations per year from its year of publication is counted in, e.g., a 2010 study would expect to have at least 45 citations\footnote{All the citations were counted by 30th September 2019.}. The only exception is for the work published in the years of writing this article (i.e., 2019), where we consider any published work or pre-press ones that have not yet been given an issue number, regardless of their citation counts. The reasons behind this setting are three-folds:

	\rev{\hspace{1em} (a) Our aim is to emphasize on the major trends about how SBSE has been used for SASs. This is important, as any issue discovered would be particularly prevalent across the most visible studies, which are of even higher impact. It, therefore, makes sense to ``sample" the literature for the most ``representative'' work. This approach was adopted by many studies, such as~\cite{DBLP:journals/infsof/FuMS16}, where they used the citation count from Google Scholar as a threshold to select studies for review, as we did in this work.}

	\rev{\hspace{1em} (b) It is not uncommon to see that software engineering surveys are conducted using some metrics to measure the ``impact" of a work. For example, some restrict their work only at what the authors believe to be premium venues~\cite{DBLP:journals/tse/GalsterWTMA14}, others use a threshold on the impact factors of the published journals, e.g., Cai and Card~\cite{cai2008analysis} used $0.55$, and Zou et al.~\cite{8466000} used $2.0$. In our case, it may not be a best practice to apply a metric at the venue level as the work on SASs often cuts across different fields (as we will show in Table~\ref{tb:papers-count}) --- it is difficult to quantify the ``impact" across communities. We, therefore, have taken a measurement at the paper level based on the citation counts from Google Scholar, which has been used as the metric to differentiate between the studies in some prior work~\cite{ten-year-sbse,DBLP:journals/tse/GalsterWTMA14,DBLP:journals/infsof/FuMS16}.}

	\rev{\hspace{1em} (c) Indeed, there is no rule to set the citation threshold. These may seem very high at the first glance, but are in fact reasonable due to two reasons: (i) by publication date, we meant the official date that the work appears on the publisher's webpage (for journal work, this means it has been given an official issue number). Yet, it is not uncommon that many studies are made citable as pre-prints before the actual publication, e.g., ICSE often has around 6 months gap between notification and official publication, and there is an even larger gap for some journals. This has helped to accumulate citations. (ii) Google Scholar counts the citations made by any publicly available documents and self-citation, which can still be part of the impact but implies their citation count may be higher than those purely made by peer-reviewed publications. Nevertheless, this could indeed pose a threat of construct validity, which we will discuss in Section~\ref{sec:tov}.}


\item The study is a short or work-in-progress paper, i.e., shorter than 8 pages (double column) or 15 pages (single column).
		
\item The study is a review, survey, tutorial, or purely empirical work.
		
\item The study is published in a non-peer-reviewed public venue, e.g., arXiv.

\end{enumerate}

Finally, if multiple studies of the same research work are found, we applied the following cleaning criteria to determine if they should all be considered. The same procedure is applied if the same authors have published different studies for the same SBSE approach, and thereby only significant contributions are analyzed for the review.

\begin{itemize}
\item All studies are considered if they report on the same problem but have different solutions.

\item All studies are considered if they report on the same problem and solutions, but have different assumptions about the nature of the problem or have new findings.

\item  When the above two criteria do not hold, only the latest version or the extended journal version is considered.

\end{itemize}

\begin{table}
\caption{Data collection items.}
\label{tb:items}
\setlength{\tabcolsep}{1mm}
\centering

\begin{tabular}{ccc}\toprule

\textbf{ID}&\textbf{Item}&\textbf{RQ}\\ 
\midrule

$I_1$&Author(s)&N/A\\
$I_2$&Year&N/A\\
$I_3$&Title&N/A\\
$I_4$&Venue (journal or conference)&N/A\\
$I_5$&Citation count&N/A\\
$I_6$&Selected search algorithm(s) and reasons&RQ1\\
$I_7$&\# algorithm(s) compared quantitatively&RQ1\\
$I_8$&SAS problem(s) to be searched&RQ1,RQ2\\

$I_{9}$&Multi-objectivity formalization and reasons&RQ2\\
$I_{10}$&Formalization assumptions&RQ2\\

$I_{11}$&Quality indicator for multiple objectives and reasons&RQ3\\

$I_{12}$&Domain information in search and reasons&RQ4\\
$I_{13}$&\makecell{Aspect(s) of specialization in search algorithm(s)}&RQ4\\

$I_{14}$&Subject SAS(s) used and reasons&RQ5\\

\bottomrule
\end{tabular}
\end{table}

\subsection{Data Items and Classification}
\label{sec:item-class}

\rev{The key items to be collected when reviewing the details of the primary studies have been shown in Table~\ref{tb:items}. We now describe their design rationales and the procedure to extract and classify the data from each item.}

\rev{The data for $I_1$ to $I_4$ is merely used as the meta-information of the primary studies. $I_5$ and $I_6$, which answer \textbf{RQ1}, aim to identify the most widely used search algorithms for SAS and the justifications of their choices. In general, the essential ways to justify the choice of a search algorithm lies in two forms: (i) it is theoretically justified by discussing why do the characteristics of search algorithm(s) align well with the requirements of the problem, with or without contrast to the applicable alternatives. This includes, e.g., how its pros and/or cons fit with the SAS problem, or how its success in other cases can be applied under the current problem; (ii) it is experimentally justified by comparing with at least one other applicable alternative algorithm in some aspects, e.g., optimality, convergence trajectory, or landscape coverage. Note that in theoretical justification, it is reasonable that a search algorithm is chosen because previous work has shown that it is the best for the SAS problem considered. In this case, however, we look for evidence or assertion to justify that the current SAS problem studied is identical (or at least share many similarities) to those from the previous work. Therefore, simply stating that a search algorithm is chosen \textit{because it has been widely used in previous work} is not a theoretical justification considered in this work. To understand whether such justifications are reasonable, we classified the data item into the following levels:}

\begin{itemize}
	\item \rev{$L_1$: Both theoretical and experimental justifications are available.}
	\item \rev{$L_2$: Only theoretical justifications are discussed.}
	\item \rev{$L_3$: Only experimental justifications are presented.}
	\item \rev{$L_4$: Neither theoretical nor experimental justifications is available.}
\end{itemize}

\rev{Clearly, $L_1$ is the most ideal situation, and $L_4$ would imply a lack of justification. We also place having theoretical justifications being more important than experimental comparison, as when justifying the algorithm choice in SBSE for SASs, the former can guide the design of the latter but rarely the other way around (we will discuss this with more details in Section~\ref{sec:rq1}). $I_8$ provides detailed information for both \textbf{RQ1} and \textbf{RQ2} as classified by the common categories of SAS problem~\cite{DBLP:journals/taas/SalehieT09,DBLP:conf/dagstuhl/LemosGMSA}. In particular, it additionally contains:}

\begin{itemize}
	\item[---] \rev{Managing or managed system.}
	\item[---] \rev{MAPE step(s) that involves search~\cite{DBLP:journals/computer/KephartC03}.}
	\item[---] \rev{Self-adaptation purpose(s), e.g., self-configuration or self-optimization~\cite{DBLP:journals/taas/SalehieT09}.}
	\item[---] \rev{Search objective(s).}
	\item[---] \rev{Search constraint(s).}
\end{itemize}


\rev{$I_9$ and $I_{10}$ are useful for \textbf{RQ2}. Specifically, in $I_9$, we recorded any reasons why a particular objective formulation of the search for SASs (e.g., single-objective, Pareto, and weighted) was chosen, or otherwise, it was marked as \textit{Unknown}. $I_{10}$ seeks to understand what treatment has been assumed as required by certain formulation. For example, how to select a final solution under Pareto search; how to set the weight vector for weighted search. $I_{11}$ provides data for \textbf{RQ3}, including the quality indicators used for Pareto search on SAS, and specifically the justifications of the generic quality indicators considered, e.g., HV~\cite{Zitzler1998} and IGD~\cite{Coello2004}. Again, we classify whether the justification is reasonable using the levels as below:}

\begin{itemize}
	\item \rev{$L_1$: The generic quality indicator is justified by referring to what quality aspects~\cite{DBLP:conf/icse/Li0Y18,DBLP:journals/csur/LiY19} they cover with respect to the preferences assumed in the SAS problem. By preferences in this work, we refer to the favored shift on the trade-off between different objectives.}
	\item \rev{$L_2$: The generic quality indicator is justified by referring to what quality aspects they cover only.}
	\item \rev{$L_3$: Neither the quality aspects covered nor the preference of the SAS problem is discussed.}
\end{itemize}

\rev{$L_1$ represents a well-justified case while $L_3$ can be questionable. Indeed, as discussed by Li et al.~\cite{DBLP:conf/icse/Li0Y18}, in SBSE each quality indicator may only cover certain quality aspect (e.g., convergence and diversity), and therefore its choice needs to be justified therein and aligned with the preferences of the SAS problem, e.g., whether one objective is naturally more preferred than the others.}

\rev{$I_{12}$ and $I_{13}$ answer \textbf{RQ4} by revealing what domain information of SAS (e.g., variation points, objectives, and model) has been used to specialize which aspect of a search algorithm, such as representation, fitness, and operator. We also collected the reason for leveraging a particular form of domain information. In particular, we classify the domain information from $I_{12}$ into two categories as proposed by Chen et al.~\cite{DBLP:journals/pieee/ChenBY20}:}

\begin{itemize}
	\item \rev{\textbf{Problem nature} refers to commonly known basic properties
and characteristics of the problem domain, such that
the search algorithms have to comply with in order to be
used appropriately. This may, for example, include the
type/range of the variables, sparsity of the values, forms of the
equality, and inequality constraints. Directly applying a standard search algorithm is often considered as exploiting only the problem’s nature without further specialization, due primarily to the
generality of these algorithms~\cite{DBLP:journals/pieee/ChenBY20}.}

   \item \rev{\textbf{Domain expertise} is represented as
or produced by typical SE/SAS methods, practices, and models involved in the engineering process. Most commonly, the SE/SAS knowledge of domain expertise is not
naturally intuitive form the problem context but can
be extracted through engineering practices, skills, and
tools, for example, design models, formatted documents, or even concepts.}
\end{itemize}

\rev{$I_{14}$ was designed for \textbf{RQ5} and it contains several additional data items:}  

\begin{itemize}
	\item[---] \rev{Type (real system, simulator or dataset) and domain.}
	\item[---]  \rev{Search space.}
	\item[---]  \rev{\# variation point.}
	\item[---]  \rev{Types of environment changes, e.g., workload, signal or service availability.}
	\item[---]  \rev{Reasons of selected subject SAS(s).}
	\item[---]  \rev{\# subject SASs (from different settings or domains).}
	\item[---]  \rev{\# subject SASs (from different domains only).}
\end{itemize}

\subsection{Data Collection Process}
\label{sec:data-collection}

\rev{For each primary study identified, the data items from Table~\ref{tb:items} were collected and classified based on the coding from Section~\ref{sec:item-class}. To this end, the first author of this paper and two other researchers acted as the investigators and reviewed the primary studies independently. The data and classification extracted by one were checked by each other. Disagreements and discrepancies were resolved by discussing among the investigators or by consulting other authors. In this work, we adopted three iterations for the data collection process following the recommendation from a recent survey~\cite{8466000}:}
		
\rev{\textit{\underline{Iteration 1:}} This iteration aims to conduct an initial data collection to summarize the data and perform preliminary classification. In particular, for those data items that do not have clearly pre-defined categories (e.g., $I_9$ and $I_{10}$), each investigator proposed his own categories without counseling each other.}

\rev{\textit{\underline{Iteration 2:}} In this iteration, all investigators checked the data and classification from each other to ensure consistency. A study was discussed during the process if there is any discrepancy in (i) the classification; (ii) the self-defined categories; (iii) the data itself. All the concerned studies and their data items were examined in order to reach an agreement. Further reading to understand the root cause of the discrepancy was conducted when necessary. Overall, 32 studies were discussed and $I_{12}$ being the data item that was involved in most of the discussions, which is perhaps due to the fact that many studies contain a mix of different forms of domain information when using SBSE for SASs.}

\rev{\textit{\underline{Iteration 3:}} The process of the final iteration is similar to that of \textit{Iteration 1}, but its goal is to eliminate any typo, missing labels, and errors.}


\section{Discussions on Results, Disappointments and Opportunities}
\label{sec:rq}

\begin{table*}[t!]
  \caption{The primary studies and their venues (sorted in descending order based on the primary studies count).}
  
\label{tb:papers-count} 
\begin{tabularx}{\textwidth}{p{8cm}P{2.5cm}P{0.5cm}X}
\toprule 
\multicolumn{1}{c}{\textbf{Journal}} & \textbf{Candidate
Studies} &\multicolumn{2}{c}{ \textbf{Primary Studies}}
\\
\midrule 

IEEE Transactions on Software Engineering & 21 & 8 &\cite{DBLP:journals/tse/CalinescuGKMT11}~\cite{DBLP:journals/tse/MalekMM12}~\cite{DBLP:journals/tse/CardelliniCGIPM12}~\cite{DBLP:journals/tse/RosaRLHS13}~\cite{DBLP:journals/tse/MoserRD12}~\cite{DBLP:journals/tse/EsfahaniEM13}~\cite{DBLP:journals/tse/NallurB13}~\cite{DBLP:journals/tse/WangHYY18} \\ \hline 
Elsevier Journal of Systems and Software & 28 & 8 &\cite{DBLP:journals/jss/VerbelenSSTD11}~\cite{DBLP:journals/jss/ChengG12}~\cite{DBLP:journals/jss/Potena13}~\cite{DBLP:journals/jss/PascualLPFE15}~\cite{DBLP:journals/jss/BashariBD18}~\cite{DBLP:journals/jss/CalinescuCGKP18}~\cite{DBLP:journals/jss/XuB19a}~\cite{DBLP:journals/jss/Gerostathopoulos19} \\ \hline
ACM Transactions on Autonomous and Adaptive Systems & 25 & 7 &\cite{DBLP:journals/taas/LewisECRTY15}~\cite{DBLP:journals/taas/Garcia-GalanPTC16}~\cite{DBLP:journals/taas/ZoghiSLG16}~\cite{DBLP:journals/taas/SuchR16}~\cite{kinneer2019information}~\cite{DBLP:journals/taas/ShevtsovWM19}~\cite{DBLP:journals/taas/FilhoP17} \\ \hline 
IEEE Transactions on Services Computing & 19 & 3 &\cite{DBLP:journals/tsc/LeitnerHD13}~\cite{DBLP:journals/tsc/SharmaR19}~\cite{DBLP:journals/tsc/ChenB17} \\ \hline 
Elsevier Future Generation Computer Systems & 31 & 3 &\cite{DBLP:journals/fgcs/VerbelenSTD13}~\cite{DBLP:journals/fgcs/PascualPF15}~\cite{DBLP:journals/fgcs/BarakatML18} \\ \hline  
Springer Automated Software Engineering & 8 & 2 &\cite{DBLP:journals/ase/GerasimouCT18}~\cite{DBLP:journals/ase/DellAnnaDD19} \\ \hline 
Elsevier Information Science & 19 & 1 &\cite{DBLP:journals/isci/NascimentoL17} \\ \hline 
IEEE Transactions on Mobile Computing & 15 & 1 &\cite{DBLP:journals/tmc/PaolaFGRD17} \\ \hline 
IEEE Transactions on Cloud Computing & 13 & 1 &\cite{7523230} \\ \hline 

Springer Service Oriented Computing and Applications & 7 & 1 &\cite{DBLP:journals/soca/HuberHKBK14} \\ \hline

Elsevier Information and Software Technology & 4 & 1 &\cite{DBLP:journals/infsof/ChenLY19} \\ \hline 
ACM Transactions on Software Engineering and Methodology & 1 & 1 &\cite{DBLP:journals/tosem/ChenLBY18} \\ \hline 
ACM Transactions on Parallel Computing & 1 & 1 &\cite{DBLP:journals/topc/BehzadBPS19} \\ \hline 
Proceedings of IEEE& 1 & 1 &\cite{DBLP:journals/pieee/TsigkanosMD19} \\ \midrule 
\multicolumn{1}{c}{\textbf{Conference and Symposium}}&\multicolumn{3}{c}{} \\
\midrule
IEEE/ACM Symposium on Software Engineering for Adaptive &42 & 12 &\cite{DBLP:conf/icse/PascualPF13}~\cite{DBLP:conf/icse/ChenB14}~\cite{DBLP:conf/icse/Garcia-GalanPTC14} ~\cite{DBLP:conf/icse/CamaraMG14}~\cite{DBLP:conf/icse/GerasimouCB14}~\cite{DBLP:conf/icse/FredericksDC14}~\cite{DBLP:conf/icse/HerbstKWG15}~\cite{DBLP:conf/icse/CailliauL17}~\cite{DBLP:conf/icse/KinneerCWGG18}\\  

and Self-Managing Systems&  &   &\cite{DBLP:conf/icse/Gerostathopoulos18}~\cite{8787163}~\cite{DBLP:conf/icse/JamshidiCSKG19}\\ \hline 

ACM Joint European Software Engineering Conference and Symposium on the Foundations of Software Engineering & 27 & 5  &\cite{DBLP:conf/sigsoft/ElkhodaryEM10}~\cite{DBLP:conf/sigsoft/EsfahaniKM11}~\cite{DBLP:conf/sigsoft/FilieriHM15}~\cite{DBLP:conf/sigsoft/ShevtsovW16}~\cite{DBLP:conf/sigsoft/MaggioPFH17}\\ \hline

IEEE Conference on Autonomic Computing & 13 & 3  &\cite{DBLP:conf/icac/RamirezKCM09}~\cite{DBLP:conf/icac/MorenoCGS16}~\cite{DBLP:conf/icac/GhahremaniG017}\\ \hline 
ACM Conference on Genetic and Evolutionary Computation & 5 & 3  &\cite{DBLP:conf/gecco/WuWHJK15}~\cite{DBLP:conf/gecco/HaraldssonWBS17}~\cite{DBLP:conf/gecco/0001LY18}\\ \hline 

IEEE Conference on Services Computing & 37 & 2  &\cite{DBLP:conf/IEEEscc/MiWYZSY10}~\cite{DBLP:conf/IEEEscc/JiangPLC12}\\ \hline 

IEEE Conference on Self-Adaptive and Self-Organizing Systems & 31 & 2  &\cite{DBLP:conf/saso/FredericksGK019}~\cite{DBLP:conf/saso/PodolskiyMKGP19}\\ \hline
IEEE/ACM Conference on Automated Software Engineering & 22 & 2  &\cite{DBLP:conf/kbse/WangM09}~\cite{DBLP:conf/kbse/GerasimouTC15}\\ \hline 

ACM Conference on Performance Engineering& 19 & 1  &\cite{DBLP:conf/wosp/0001BWY18}\\ \hline

IEEE/ACM Conference on Software Engineering & 17 & 1  &\cite{DBLP:conf/icse/ChenPYNZ14}\\ \hline 

IEEE Conference on Requirements Engineering& 9 & 1  &\cite{DBLP:conf/re/PengCYZ10}\\ \hline 

Springer Conference on Fundamental Approaches to Software Engineering& 6 & 1  &\cite{DBLP:conf/fase/CalinescuGB15}\\ \hline

IEEE Conference on Software Architecture & 3 & 1  &\cite{DBLP:conf/icsa/CalinescuCGKP17}\\ \hline 
ACM Conference on Systems and Software Product Line & 2 & 1  &\cite{DBLP:conf/splc/WeckesserKPMSB18}\\   
  
  \midrule  

\multicolumn{1}{c}{\textbf{Total}} & 409 & 74 \\ \bottomrule  
\end{tabularx}
\end{table*}

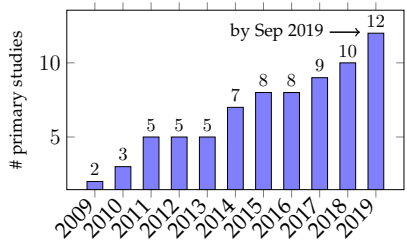
\begin{figure}
  \centering
\begin{tikzpicture}
\begin{axis}[
    ybar=0.05cm,
    width=8cm,
    height=5cm,
    ymax=13,
    bar width=0.3cm,
 enlarge x limits=0.1,
 enlarge y limits=0.05,
    legend style={at={(0.1,0.96)},
      anchor=north,legend columns=2},
    ylabel={\# primary studies},
     nodes near coords,
    symbolic x coords={2009,2010,2011,2012,2013,2014,2015,2016,2017,2018,2019},
    y tick label style={font=\large},
     x tick label style={rotate=45,anchor=east,font=\large},
    xtick=data
    ]
\addplot[fill=blue!50] coordinates {(2009,2) (2010,3) (2011,5) (2012,5) (2013,5) (2014,7) (2015,8) (2016,8) (2017,9) (2018,10) (2019,12)};



\end{axis}

 \node[align=left,anchor=west,black] (source) at (3,3){by Sep 2019};
   \node (destination) at (5.7,3){};
    \draw[->][black,thick](source)--(destination);
\end{tikzpicture}
    \caption{Number of primary studies identified per year.}
  \label{fig:paper-count}

  \end{figure}

To provide an overview, we show all the 74 primary studies and their published venues in Table~\ref{tb:papers-count}, based on which it is clear that the studies identified come from a variety of well-known conferences and journals\footnote{We omitted the venues that do not result in any primary study.}. Figure~\ref{fig:paper-count} illustrates the evolution of study count with respect to the publication year. We note that the number of studies increases at a steady pace, achieving a 5$\times$ increment in 2019 (by September) compared with 2009. This implies an increasing popularity of the research in SBSE for SASs. 

It is worth noting that the primary studies do not only contain work published in top Software Engineering venues, but also those relevant ones that were published in System Engineering conferences/journals as well as those in
Computational Optimization venues, as long as they are related to problems in engineering SAS and comply with the inclusive/exclusive criteria. This evidences the fact that SBSE for SASs work is often interdisciplinary, spanning across different communities.

In what follows, we present and analyze the survey results with respect to our RQs, together with the disappointments, the justifications of the likely issues and opportunities to mitigate them. All the data of our survey and experiments is publicly available at our repository\footnote{\url{https://github.com/taochen/sbse-for-sas}}. 

\begin{figure}[!t]
  \centering
  \begin{subfigure}[t]{0.515\columnwidth}
\begin{tikzpicture}
\begin{axis}[
    ybar stacked,
	bar width=10pt,
	ymax=15,
	nodes near coords, 
	nodes near coords={\pgfmathfloatifflags{\pgfplotspointmeta}{0}{}{\pgfmathprintnumber{\pgfplotspointmeta}}},
    enlargelimits=0,
    enlarge x limits=0.05,
    legend style={at={(0.65,0.99),font=\large},
      anchor=north,legend columns=-1},
    ylabel={\# study-algorithm pairs},
    xtick={2009,2010,...,2019},
    y label style={at={(-0.08,0.5)},font=\large},
    symbolic x coords={2009, 2010, 2011, 2012, 
		2013, 2014, 2015, 2016, 2017, 2018, 2019},
       x tick label style={rotate=45,anchor=east,font=\large},
    y tick label style={font=\large}
    ]
\addplot[ybar, fill=white, draw=black] plot coordinates {(2009,0) (2010,0) 
  (2011,6) (2012,0) (2013,0) (2014,0) (2015,0) (2016,5) (2017,0) (2018,1) (2019,1)};
  
  \addplot[ybar, fill=blue!15, draw=black] plot coordinates {(2009,0) (2010,1) 
  (2011,0) (2012,1) (2013,1) (2014,1) (2015,1) (2016,0) (2017,0) (2018,1) (2019,0)};
  
    \addplot[ybar, fill=blue!35, draw=black] plot coordinates {(2009,0) (2010,0) 
  (2011,2) (2012,0) (2013,2) (2014,0) (2015,2) (2016,2) (2017,2) (2018,1) (2019,4)};
  
    \addplot[ybar, fill=blue!55, draw=black] plot coordinates {(2009,2) (2010,2) 
  (2011,3) (2012,4) (2013,2) (2014,5) (2015,3) (2016,2) (2017,5) (2018,4) (2019,7)};

\legend{$L_1$, $L_2$, $L_3$, $L_4$}

\end{axis}
\coordinate (A) at (0.35,1.5);
\coordinate (B) at (0.95,2.6);
\coordinate (C) at (1.55,5);
\coordinate (D) at (2.15,2.2);
\coordinate (E) at (2.8,2.7);
\coordinate (F) at (3.4,2.6);
\coordinate (G) at (4,2.7);
\coordinate (H) at (4.7,3.8);
\coordinate (I) at (5.3,2.6);
\coordinate (J) at (5.9,3);
\coordinate (K) at (6.55,4.9);

\node[rotate=90] at (A) {GA/ES};
\node[rotate=90] at (B) {GA/ES/IP solver};
\node[rotate=90] at (C) {IP solver};
\node[rotate=90] at (D) {ES};
\node[rotate=90] at (E) {IP solver};
\node[rotate=90] at (F) {ES};
\node[rotate=90] at (G) {GA};
\node[rotate=90] at (H) {GA};
\node[rotate=90] at (I) {ES};
\node[rotate=90] at (J) {GA};
\node[rotate=90] at (K) {ES};
\end{tikzpicture}
    \subcaption{Single/aggregated objective search}
 \label{fig:alg-s-evo}
    \end{subfigure}
    \hspace{-0.2cm}
      \begin{subfigure}[t]{0.485\columnwidth}
\begin{tikzpicture}
\begin{axis}[
    ybar stacked,
	bar width=10pt,
	ymax=15,
	nodes near coords, 
	nodes near coords={\pgfmathfloatifflags{\pgfplotspointmeta}{0}{}{\pgfmathprintnumber{\pgfplotspointmeta}}},
    enlargelimits=0,
     enlarge x limits=0.5,
    legend style={at={(0.65,0.99),font=\large},
      anchor=north,legend columns=-1},
    symbolic x coords={2014, 2015, 2016, 2017, 2018, 2019},
    xtick=data,
    x tick label style={rotate=45,anchor=east,font=\large},
    y tick label style={font=\large}
    ]
\addplot[ybar, fill=white, draw=black] plot coordinates {(2014,0) (2015,0) (2016,0) (2017,0) (2018,0) (2019,0)};
  
  \addplot[ybar, fill=blue!15, draw=black] plot coordinates {(2014,0) (2015,0) (2016,0) (2017,0) (2018,1) (2019,1)};
  
    \addplot[ybar, fill=blue!35, draw=black] plot coordinates {(2014,0) (2015,10) (2016,2) (2017,2) (2018,7) (2019,4)};
  
    \addplot[ybar, fill=blue!55, draw=black] plot coordinates {(2014,1) (2015,0) (2016,1) (2017,0) (2018,1) (2019,0)};

\legend{$L_1$, $L_2$, $L_3$, $L_4$}

\end{axis}
\coordinate (A) at (1.7,1.2);
\coordinate (B) at (2.4,4.6);
\coordinate (C) at (3.15,2.7);
\coordinate (D) at (3.8,2.3);
\coordinate (E) at (4.45,4.2);
\coordinate (F) at (5.15,2.7);
\node[rotate=90] at (A) {FastPGA};
\node[rotate=90] at (B) {NSGA-II};
\node[rotate=90] at (C) {NSGA-II/FastPGA};
\node[rotate=90] at (D) {NSGA-II/MOACO};
\node[rotate=90] at (E) {NSGA-II};
\node[rotate=90] at (F) {NSGA-II};
\end{tikzpicture}
    \subcaption{Pareto search}
 \label{fig:alg-m-evo}
    \end{subfigure}
    \caption{Popularity evolution of search algorithm and their levels of justification for SAS (12 studies use more than one algorithm).}
     \label{fig:alg-evo}
  \end{figure}
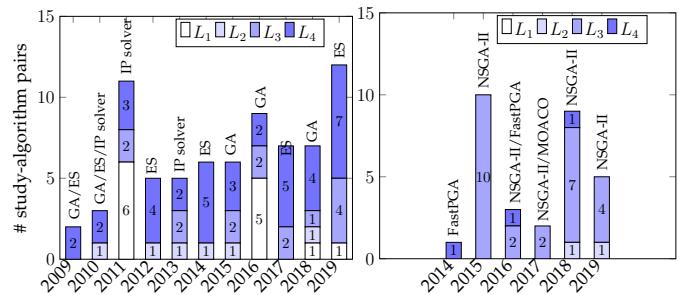
  
\begin{table*}[t!]
\caption{Context in which the top 10 search algorithms are used for SAS.}
\label{tb:alg-context}
\setlength{\tabcolsep}{1.88mm}
\centering

\begin{tabular}{cccccc}\toprule

\textbf{Algorithm}\hfill&\textbf{Problem}&\textbf{Managed(ing)}&\textbf{MAPE-K phase}&\textbf{Self-adaptation purpose}&\textbf{\# Objectives}\\ 
\midrule

ES&\makecell{Configuration (13),\\Deployment (2), \\Prediction (1),\\Requirement (1)}&\makecell{Managing (13),\\Managed (1)}&\makecell{Plan (14),\\Analyze (2)}&\makecell{Self-optimization (8), Self-configuration (14),\\Self-healing (2)}&-\\

\hline

GA&\makecell{Configuration (12),\\Deployment (2)}&Managing (13)&Plan (13)&\makecell{Self-optimization (13), Self-configuration (13)}&-\\

\hline

IP Solver&\makecell{Configuration (8),\\Deployment (2)}&Managing (10)&Plan (10)&\makecell{Self-optimization (10), Self-configuration (9)}&-\\

\hline

RS&\makecell{Configuration (3)}&Managing (3)&Plan (3)&\makecell{Self-optimization (2), Self-configuration (3)}&-\\

\hline

GP&\makecell{Deployment (2),\\Code improvement (1)}&\makecell{Managing (3))}&\makecell{Plan (3)}&\makecell{Self-optimization (2), Self-configuration (2),\\Self-healing (1)}&-\\

\hline

GS&\makecell{Configuration (1), \\Deployment (1)}&Managing (2)&Plan (2)&\makecell{Self-optimization (2), Self-configuration (1)}&-\\

\hline

BB&\makecell{Configuration (2)}&Managing (2)&Plan (2)&\makecell{Self-optimization (2), Self-configuration (2)}&-\\

\hline

Simplex&\makecell{Configuration (2)}&Managing (2)&Plan (2)&\makecell{Self-optimization (2), Self-configuration (2)}&-\\

\hline

BOGP&\makecell{Configuration (2)}&Managing (2)&Plan (2)&\makecell{Self-optimization (2), Self-configuration (2)}&-\\

\hline

DP&\makecell{Configuration (2)}&Managing (2)&Plan (2)&\makecell{Self-optimization (2), Self-configuration (2)}&-\\

\hline
\hline

NSGA-II&\makecell{Configuration (10),\\Code improvement (1)}&\makecell{Managing (10),\\Managed (1)}&Plan (11)&\makecell{Self-optimization (11), Self-configuration (11)}&\makecell{3 (5), 2 (5),\\3-6 (1)}\\

\hline

SPEA2&\makecell{Configuration (3),\\Deployment (2)}&\makecell{Managing (5)}&Plan (5)&\makecell{Self-optimization (5), Self-configuration (5)}&3 (5)\\

\hline

MOCell&\makecell{Configuration (3)}&\makecell{Managing (3)}&Plan (3)&\makecell{Self-optimization (3), Self-configuration (3)}&3 (3)\\

\hline

IBEA&\makecell{Configuration (3)}&\makecell{Managing (3)}&Plan (3)&\makecell{Self-optimization (3), Self-configuration (3)}&3 (2), 2 (1)\\

\hline

FastPGA&\makecell{Configuration (2)}&\makecell{Managing (2)}&Plan (2)&\makecell{Self-optimization (2), Self-configuration (2)}&3-6 (2)\\

\hline

MOEA/D-STM&\makecell{Configuration (2)}&\makecell{Managing (2)}&Plan (2)&\makecell{Self-optimization (2), Self-configuration (2)}&2 (2)\\

\hline

MOEA/D&\makecell{Configuration (1)}&\makecell{Managing (1)}&Plan (1)&\makecell{Self-optimization (1), Self-configuration (1)}&3 (1)\\

\hline

MOACO&\makecell{Configuration (1)}&\makecell{Managing (1)}&Plan (1)&\makecell{Self-optimization (1), Self-configuration (1)}&5 (1)\\

\hline

MOCHC&\makecell{Configuration (1)}&\makecell{Managing (1)}&Plan (1)&\makecell{Self-optimization (1), Self-configuration (1)}&3 (1)\\

\hline

PAES&\makecell{Configuration (1)}&\makecell{Managing (1)}&Plan (1)&\makecell{Self-optimization (1), Self-configuration (1)}&3 (1)\\

\bottomrule
\end{tabular}
\begin{tablenotes}
    \footnotesize
    \item Number in the bracket indicates how many studies are involved.
\end{tablenotes}
\end{table*}

\subsection{RQ1: Search Algorithms for SASs}
\label{sec:rq1}

\subsubsection{Significance} 

Being one of the most important parts of SBSE, understanding what, where, and why search algorithms are chosen in SBSE for SASs is essential. 


\subsubsection{Findings} 

From Figure~\ref{fig:alg-evo}, we can clearly see that the most popular search algorithms used each year have been similar, i.e., Exhaustive Search (ES)~\cite{DBLP:journals/tse/CalinescuGKMT11,DBLP:conf/icse/CailliauL17}, Genetic Algorithm (GA)~\cite{DBLP:conf/icac/RamirezKCM09,DBLP:journals/ase/GerasimouCT18} and Integer Programming (IP) solver\footnote{We have seen that a variety of solvers used, e.g., CPLEX (\url{https://www.ibm.com/analytics/cplex-optimizer}), LINDO (\url{https://www.lindo.com/}) and SCIP (\url{https://scip.zib.de/})~\cite{DBLP:journals/tse/EsfahaniEM13,DBLP:journals/tse/WangHYY18}, in SBSE for SASs.} for single/aggregated objective search while NSGA-II~\cite{DBLP:journals/tec/DebAPM02} for Pareto search in SAS~\cite{DBLP:journals/tosem/ChenLBY18,DBLP:journals/jss/CalinescuCGKP18}. A more important message we obtain from the results is that $L_4$ (e.g.,~\cite{DBLP:conf/icac/RamirezKCM09,DBLP:journals/soca/HuberHKBK14,DBLP:journals/isci/NascimentoL17}) and $L_3$ (e.g.,~\cite{DBLP:journals/jss/PascualLPFE15,DBLP:journals/taas/Garcia-GalanPTC16,DBLP:journals/tsc/ChenB17}) are the most common level of justification on the algorithm choice for the single/aggregated objective search and Pareto search cases, respectively. This trend does not seem to have the tendency to change according to the evolution over years. In addition to the studies where more than one search algorithm has been experimentally compared (but only one is chosen), we also found 12 studies (e.g.,~\cite{DBLP:conf/icse/Garcia-GalanPTC14,DBLP:journals/jss/PascualLPFE15,DBLP:journals/tosem/ChenLBY18}) which have chosen multiple search algorithms, because the proposed approach is algorithm agnostic.

  \begin{figure}[!t]
  \centering
\begin{tikzpicture}
\begin{axis}[
 xbar stacked,
    width=7cm,
    height=5cm,
    bar width=0.25cm,
    	nodes near coords, 
	nodes near coords={\pgfmathfloatifflags{\pgfplotspointmeta}{0}{}{\pgfmathprintnumber{\pgfplotspointmeta}}},
  enlargelimits=0,
 enlarge y limits=0.05,
    legend style={at={(0.85,0.65)},
      anchor=north,legend columns=1},
    xlabel={\# primary studies},
    symbolic y coords={Dynamic Programming {(DP)},Bayesian Optimization {(BO)},Simplex {(Simplex)},Branch and Bound {(BB)},Greedy Search {(GS)},Genetic Programming {(GP)},Random Search {(RS)},Mixed Integer {(Nonlinear)} Programming {(MIP)} Solver,Genetic Algorithm {(GA)},Exhaustive Search {(ES)}},
     yticklabels={Exhaustive Search \textbf{(ES)},Genetic Algorithm \textbf{(GA)},(Mixed) Integer (Nonlinear) Programming \textbf{(IP)} Solver,Random Search \textbf{(RS)},Genetic Programming \textbf{(GP)},Greedy Search \textbf{(GS)},Branch and Bound \textbf{(BB)},Simplex \textbf{(Simplex)},Bayesian Optimization Gaussian Process \textbf{(BOGP)},Dynamic Programming \textbf{(DP)}    
     },
         ytick=data
    ]
\addplot[xbar, fill=white, draw=black] coordinates {(0,Exhaustive Search {(ES)}) (3,Genetic Algorithm {(GA)}) (2,Mixed Integer {(Nonlinear)} Programming {(MIP)} Solver) (0,Random Search {(RS)}) (2,Genetic Programming {(GP)}) (2,Greedy Search {(GS)}) (1,Branch and Bound {(BB)}) (0,Simplex {(Simplex)}) (0,Bayesian Optimization {(BO)}) (0,Dynamic Programming {(DP)})};

\addplot[xbar, fill=blue!15, draw=black] coordinates {(3,Exhaustive Search {(ES)}) (0,Genetic Algorithm {(GA)}) (1,Mixed Integer {(Nonlinear)} Programming {(MIP)} Solver) (0,Random Search {(RS)}) (0,Genetic Programming {(GP)}) (0,Greedy Search {(GS)}) (0,Branch and Bound {(BB)}) (0,Simplex {(Simplex)}) (1,Bayesian Optimization {(BO)}) (0,Dynamic Programming {(DP)})};

\addplot[xbar, fill=blue!35, draw=black] coordinates {(3,Exhaustive Search {(ES)}) (4,Genetic Algorithm {(GA)}) (0,Mixed Integer {(Nonlinear)} Programming {(MIP)} Solver) (2,Random Search {(RS)}) (0,Genetic Programming {(GP)}) (0,Greedy Search {(GS)}) (1,Branch and Bound {(BB)}) (0,Simplex {(Simplex)}) (1,Bayesian Optimization {(BO)}) (0,Dynamic Programming {(DP)})};

\addplot[xbar, fill=blue!55, draw=black] coordinates {(9,Exhaustive Search {(ES)}) (6,Genetic Algorithm {(GA)}) (7,Mixed Integer {(Nonlinear)} Programming {(MIP)} Solver) (1,Random Search {(RS)}) (1,Genetic Programming {(GP)}) (0,Greedy Search {(GS)}) (0,Branch and Bound {(BB)}) (2,Simplex {(Simplex)}) (0,Bayesian Optimization {(BO)}) (2,Dynamic Programming {(DP)})};
\legend{$L_1$, $L_2$, $L_3$, $L_4$}
\end{axis}
\end{tikzpicture}
    \caption{Top 10 selected single and aggregated objective search algorithms and their levels of justification in SBSE for SASs over years.}
 \label{fig:alg-count1}
  \end{figure}
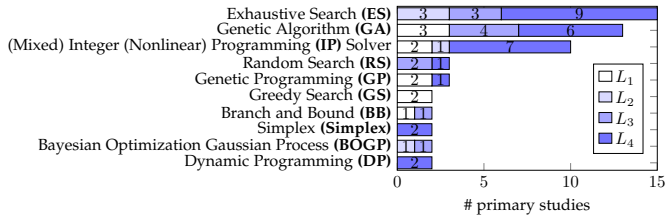

\begin{figure}[!t]
  \centering
\begin{tikzpicture}
\begin{axis}[
 xbar stacked,
    width=7cm,
    height=5cm,
    bar width=0.25cm,
	nodes near coords, 
	nodes near coords={\pgfmathfloatifflags{\pgfplotspointmeta}{0}{}{\pgfmathprintnumber{\pgfplotspointmeta}}},
  enlargelimits=0,
 enlarge y limits=0.05,
      legend style={at={(0.85,0.65)},
      anchor=north,legend columns=1},
     xlabel={\# primary studies},
    symbolic y coords={Pareto Archived Evolution Strategy {(PAES)},Multi-Objective CHC {(MOCHC)},Ant Colony Optimization {(ACO)},MOEA/D,MOEA/D-Stable Matching {(MOEA/D-STM)},Fast Pareto Genetic Algorithm {(FastPGA)},Indicator Based Evolutionary Algorithm {(IBEA)},Multi-Objective Cellular Genetic Algorithm {(MOCell)},Strength Pareto Evolutionary Algorithm2 {(SPEA2)},Non-dominated Sorting Genetic Algorithm-II {(NSGA-II)}},
           yticklabels={Non-dominated Sorting Genetic Algorithm-II \textbf{(NSGA-II)}~\cite{DBLP:journals/tec/DebAPM02},Strength Pareto Evolutionary Algorithm2 \textbf{(SPEA2)}~\cite{Zitzler2002},Multi-Objective Cellular Genetic Algorithm \textbf{(MOCell)}~\cite{nebro2009mocell},Indicator Based Evolutionary Algorithm \textbf{(IBEA)}~\cite{Zitzler2004},Fast Pareto Genetic Algorithm \textbf{(FastPGA)}~\cite{eskandari2008fast},MOEA/D-Stable Matching \textbf{(MOEA/D-STM)}~\cite{DBLP:journals/tec/LiZKLW14},Multi-Objective Ant Colony Optimization \textbf{(MOACO)}~\cite{dorigo1999ant},\textbf{MOEA/D}~\cite{Zhang2007},Multi-Objective CHC \textbf{(MOCHC)}~\cite{nebro2007optimal},Pareto Archived Evolution Strategy \textbf{(PAES)}~\cite{knowles1999pareto}          
           },
                  ytick=data
    ]
\addplot[xbar, fill=white, draw=black] coordinates {(0,Non-dominated Sorting Genetic Algorithm-II {(NSGA-II)}) (0,Strength Pareto Evolutionary Algorithm2 {(SPEA2)}) (0,Multi-Objective Cellular Genetic Algorithm {(MOCell)}) (0,Indicator Based Evolutionary Algorithm {(IBEA)}) (0,Fast Pareto Genetic Algorithm {(FastPGA)}) (0,MOEA/D-Stable Matching {(MOEA/D-STM)}) (0,Ant Colony Optimization {(ACO)}) (0,MOEA/D) (0,Multi-Objective CHC {(MOCHC)}) (0,Pareto Archived Evolution Strategy {(PAES)})};

\addplot[xbar, fill=blue!15, draw=black] coordinates {(0,Non-dominated Sorting Genetic Algorithm-II {(NSGA-II)}) (2,Strength Pareto Evolutionary Algorithm2 {(SPEA2)}) (0,Multi-Objective Cellular Genetic Algorithm {(MOCell)}) (0,Indicator Based Evolutionary Algorithm {(IBEA)}) (0,Fast Pareto Genetic Algorithm {(FastPGA)}) (0,MOEA/D-Stable Matching {(MOEA/D-STM)}) (0,Ant Colony Optimization {(ACO)}) (0,MOEA/D) (0,Multi-Objective CHC {(MOCHC)}) (0,Pareto Archived Evolution Strategy {(PAES)})};

\addplot[xbar, fill=blue!35, draw=black] coordinates {(10,Non-dominated Sorting Genetic Algorithm-II {(NSGA-II)}) (3,Strength Pareto Evolutionary Algorithm2 {(SPEA2)}) (3,Multi-Objective Cellular Genetic Algorithm {(MOCell)}) (3,Indicator Based Evolutionary Algorithm {(IBEA)}) (1,Fast Pareto Genetic Algorithm {(FastPGA)}) (1,MOEA/D-Stable Matching {(MOEA/D-STM)}) (1,Ant Colony Optimization {(ACO)}) (1,MOEA/D) (1,Multi-Objective CHC {(MOCHC)}) (1,Pareto Archived Evolution Strategy {(PAES)})};

\addplot[xbar, fill=blue!55, draw=black] coordinates {(1,Non-dominated Sorting Genetic Algorithm-II {(NSGA-II)}) (0,Strength Pareto Evolutionary Algorithm2 {(SPEA2)}) (0,Multi-Objective Cellular Genetic Algorithm {(MOCell)}) (0,Indicator Based Evolutionary Algorithm {(IBEA)}) (1,Fast Pareto Genetic Algorithm {(FastPGA)}) (1,MOEA/D-Stable Matching {(MOEA/D-STM)}) (0,Ant Colony Optimization {(ACO)}) (0,MOEA/D) (0,Multi-Objective CHC {(MOCHC)}) (0,Pareto Archived Evolution Strategy {(PAES)})};
\legend{$L_1$, $L_2$, $L_3$, $L_4$}
\end{axis}
\end{tikzpicture}
    \caption{Top 10 selected Pareto search algorithms and their levels of justification in SBSE for SASs over years (MOEA/D stands for Multi-Objective Evolutionary Algorithm based on Decomposition).}
 \label{fig:alg-count2}
  \end{figure}
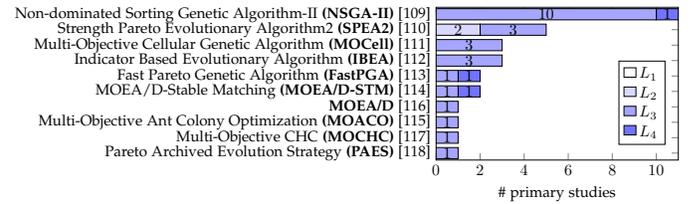

However, it is yet clear whether the overall levels of justification are biased by one or two particular search algorithms. To understand such, Figures~\ref{fig:alg-count1} and~\ref{fig:alg-count2} illustrate more clear views on the top 10 most popularly used search algorithms and their justifications of choice. For the single/aggregated objective search case, a variety of algorithms have been chosen, ranging from the exact search, e.g., ES and IP solver and the stochastic search, e.g., GA and Random Search (RS). Apparently, ES, GA, and IP solver share similar popularity but are more predominant than the rest. In the Pareto search case, NSGA-II is significantly more popular than the others --- an inherited trend from SBSE~\cite{DBLP:journals/csur/HarmanMZ12,Sayyad2013b}. In particular, we confirm that the observation of $L_3$ or $L_4$ being the most common justification level for choosing the search algorithms was not biased by a particular algorithm, but a prevalent phenomenon across all. 

Table~\ref{tb:alg-context} shows in what context the top 10 search algorithms have been used. They have clearly spanned across different SAS problems, parts of the SAS, MAPE-K phases, and self-adaptation purpose. However, in general, we see a clear trend on which some aspects are overwhelmingly targeted: the SAS configuration problems; on the managing system; at the \textit{Plan} phase and for self-optimization/-configuration purpose. In particular, for Pareto search, searching for three objectives is the most common case but NSGA-II has been applied on up to six objectives~\cite{DBLP:conf/icse/Garcia-GalanPTC14}.

Overall, our findings for \textbf{RQ1} conclude that:

\begin{tcolorbox}[breakable,left=5pt,right=5pt,top=5pt,bottom=5pt] 
\textbf{Findings 1:} In SBSE for SASs, ES, GA and IP solvers are the top three most popular algorithms for single/aggregated objective search while NSGA-II is the predominant one for Pareto search. \\
\textbf{Findings 2:} $L_3$ and $L_4$ are the most common levels of justification when choosing a search algorithm.\\ 
\textbf{Findings 3:} SBSE for SASs has been used on different contexts where it is most common to search on the managing system at \textit{Plan} phase for self-optimizing/-configuring in the SAS configuration problem.
\end{tcolorbox}

\subsubsection{Disappointments}
\label{sec:rq1-di}

Indeed, certain search algorithms are ``better known" than some others, but such a large bias is what we did not expect. In fact, it is disappointed to see that, on choosing search algorithms for SASs, majority of the studies give no justification at all ($L_4$, especially on the single/aggregated objective search) or rely on purely experimental comparisons ($L_3$, most commonly in Pareto search). We have also shown that such a finding was neither biased by a particular search algorithm nor the year of study, but overwhelmingly happened to most algorithms used in the field over the past decade --- as shown in Figure~\ref{fig:alg-evo}, out of the 97 cases, only 12 and 9 cases in total qualify to $L_1$ and $L_2$, respectively. To give a good example of $L_1$, Kinneer et al.~\cite{DBLP:conf/icse/KinneerCWGG18} state that GP is chosen because the SAS problem studied has a large search space and complex landscape and, at the same time, sub-optimal result or premature convergence is acceptable. This fits precisely with the major pros and cons of GP, and it is then supported by experimental comparisons with an alternative algorithm, i.e., ES.

Admittedly, the aim of studies in SBSE for SASs may not be finding the ``best" search algorithm for the problem. However, in whichever case, our conjecture is that the choice of search algorithm should be justifiable, i.e., ideally at $L_1$ or at least at $L_2$ if resources are rather limited, but definitely not $L_4$, because it is known that every search algorithm does have their own ``comfort zone''~\cite{DBLP:journals/tec/DolpertM97}. In fact, among those $L_3$ and $L_4$ cases, a considerable number of studies we found tend to work by analogy, i.e., one of the most common reasons that is solely used to choose an algorithm we found is \textit{``it has been widely used before or in other problems"}, e.g.,~\cite{DBLP:journals/tse/CardelliniCGIPM12,DBLP:conf/icse/Garcia-GalanPTC14,DBLP:journals/ase/GerasimouCT18}; or mostly no reasons mentioned (nor experiments) at all~\cite{DBLP:journals/isci/NascimentoL17,DBLP:journals/fgcs/BarakatML18,DBLP:journals/ase/DellAnnaDD19}. Indeed, it makes sense that a search algorithm is chosen because previous work has shown that it is the best for the SAS problem considered. In this case, however, evidence is required to justify that the current SAS problem studied is identical (or at least share many similarities) to those from the previous work. This is what we did not find in the primary studies under these cases.


All the above suggests a lack of sufficient justification on the choice of search algorithms for SAS. This implies a high risk of not being fully aware of their suitability for the SAS(s) studied and its problem, resulting in an immediate threat to the conclusion drawn. For example, ES (or similar exact search algorithms) is apparently not workable on large scale SASs~\cite{DBLP:journals/tosem/ChenLBY18}; GA may not be suitable for time-critical SASs~\cite{DBLP:journals/tsc/LeitnerHD13}; NSGA-II typically does not scale well on SAS problems with four and more objectives~\cite{Purshouse2007}. Even if the proposed approach is algorithm agnostic, a limited justification ($L_3$ or $L_4$) can still cause misleading conclusions, as we will show. Therefore, justifiably selecting an algorithm suitable for the considered SAS problem is crucial, 
which demands a well understanding of both the problem and the algorithm.

Our first disappointment is thus:

\begin{tcolorbox}[breakable,left=5pt,right=5pt,top=5pt,bottom=5pt] 
\textbf{Disappointment 1:} Unjustified bias on the choice of search algorithms. 
\end{tcolorbox}

\begin{figure}[t!]
\centering
\includegraphics[width=0.6\columnwidth]{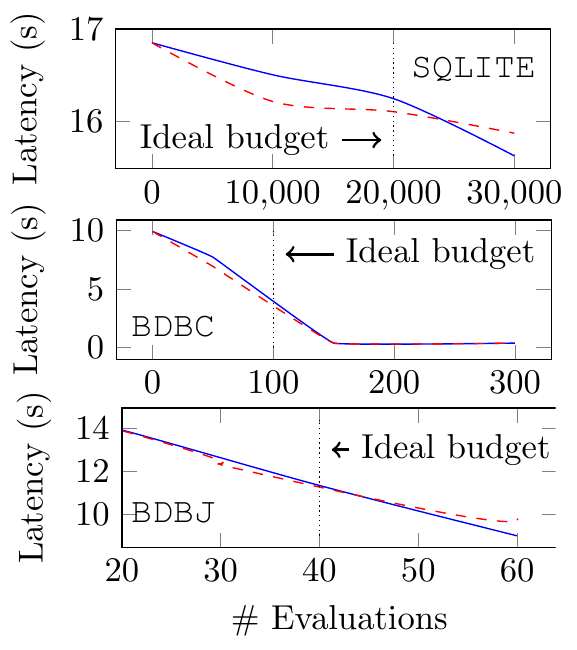}
\caption{The convergence of latency (s) by Hill Climbing (HC denoted as \textcolor{blue}{---}) and Random Search (RS denoted as \textcolor{red}{- - -}) on the three SASs from distinct domains under a workload change (each point is the mean over 100 runs). This is an example of the possible consequences when the theoretical justification is not supported by experimental justification, where the theory may be affected by misconsidered factors (e.g., search budget) of the problem.}
\label{fig:rq1-no-exp}
\end{figure}

\subsubsection{Justification on the Likely Issues}

To justify the possible issues raised form $L_2$, we first compare Hill Climbing (HC) and Random Search (RS) (as they are used in the studies and exhibit distinct characteristics) on profiling directly for three SASs, namely \textsc{SQLite}, \textsc{BDBC} and \textsc{BDBJ}, because using three different subjects, even from the same domain, can cover wider scenarios than 65\% existing work as we will show in Section~\ref{sec:rq5}. These SASs are chosen because (i) they are real-world software that has been used by prior work~\cite{DBLP:conf/icse/SiegmundKKABRS12} and (ii) they are expensive to evaluate and difficult to be thoroughly modeled, and thus design-time profiling beforehand can provide important insights on designing the policies for runtime self-adaptation. The aim is to tune latency by adjusting various variation points, e.g., the \texttt{SQLITE\_OMIT\_BTREECOUNT} on \textsc{SQLite}, under a workload change. The corresponding adaptation tactic under such a workload condition can then be drawn. Both HC and RS are run using the same search budget and their setting details, together with details of the SASs, can be found in the supplementary.



In general, one may theoretically justify that HC tends to be suitable for such a SAS problem (or better than the common baseline RS~\cite{DBLP:conf/icse/ArcuriB11}) because the HC is explicitly guided and it is known that the SASs do not contain difficult local optimum points. As we show in Figure~\ref{fig:rq1-no-exp} (each point is the mean over 100 runs), we see that such a theoretical justification is indeed the case: given sufficient search budget, HC can converge well and would eventually be better than RS. However, as can be seen from the experimental results, when both HC and RS tend to converge prematurely under a restricted budget, the RS can actually be a better fit for the problem. This implies that the above theoretical justification could misconsider the fact that the possible search budget may not be ``sufficient" enough to allow HC to converge better --- a highly likely case for the SAS problem as the profiling process can be expensive, e.g., it could take minutes to evaluate only a single solution. Now, suppose that the requirement threshold is right at the performance gap between HC and RS, then it is likely that one would mark ``no satisfactory solution" under the workload, and encode this as part of the adaptation policies. This would, of course, not be ideal as a satisfactory solution could have been found if the RS is simply used instead. The above is a typical case where $L_2$ may still fail to justify the choice of the search algorithm, due to the lack of experimental comparison that may reveal misconsidered factors in SBSE for SASs. Yet, another example is from Leitner et al.~\cite{DBLP:journals/tsc/LeitnerHD13}, who chose GA as a candidate because it is theoretically understandable that GA is less sensitive to local optimum than local search, and hence can potentially lead to better results. However, in their experiment comparison, the GA is, in fact, inferior to local search, because the local optimum points for the SAS problem are not difficult to escape from.

Next, we showcase the likely issues for $L_3$ by running NSGA-II and IBEA (as they are two most common search algorithms with distinct ``comfort zone") on synthetic service systems for runtime self-adaptation to a service change using the QWS data~\cite{DBLP:conf/smc/Al-MasriM09a}, because it is a type of the most widely used SASs as we will show in Section~\ref{sec:rq5}. We experiment on three workflows, each with a different structure and number of abstract services, ranging from 10 to 15 as used in~\cite{DBLP:journals/infsof/ChenLY19}. Again, using three different subjects (from the same domain) has already achieved better coverage than 65\% of existing studies, as what will be shown in Section~\ref{sec:rq5}. The aim is to self-adapt the SAS by re-composing the concrete services with an aim to tune different objectives upon service quality/availability changes. HV is chosen as the quality indicator because we aim to assess the overall quality of the solution set produced under no specific preferences and it covers all quality aspects of a solution set~\cite{DBLP:conf/icse/Li0Y18}. To achieve efficient search, the fitness is evaluated by using a well-defined analytical model~\cite{DBLP:conf/gecco/0001LY18,DBLP:journals/infsof/ChenLY19}. We run on three and five objective cases, as these are what have been used on NSGA-II from Table~\ref{tb:alg-context}. Both NSGA-II and IBEA use identical search budget, their setting details, and specifications of the service-based SASs can be found in the supplementary.

 \begin{table}[t!]
 
\centering
	\caption{Comparing 30 runs' mean HV for the cases of 3 (latency, throughput and availability) and 5 objectives (latency, throughput, availability, compliance and practices) by NSGA-II and IBEA on different SASs upon a service quality/availability change. This is an example of the possible consequences when the experimental justification of search algorithm is not guided by theoretical justification, where misleading conclusion may be drawn based on only the results under the three objectives.}
\label{tb:mo-rq1}

\centering
\begin{center}
\begin{tabular}{ccc|cc}\toprule

\multirow{2}{*}{\textbf{SASs}}&\multicolumn{2}{c|}{\textbf{3 Objectives}}&\multicolumn{2}{c}{\textbf{5 Objectives}}\\
\cmidrule{2-5}

&\textbf{NSGA-II}&\textbf{IBEA}&\textbf{NSGA-II}&\textbf{IBEA}\\ 
\midrule

\textsc{QWS-1}&\cellcolor{yellow!50}0.9772&0.9919&0.8683&0.9442\cellcolor{yellow!50}\\

\textsc{QWS-2}&\cellcolor{yellow!50}0.9938&0.9909&0.8959&\cellcolor{yellow!50}0.9378\\

\textsc{QWS-3}&\cellcolor{yellow!50}0.9908&0.9875&0.8659&\cellcolor{yellow!50}0.9341\\

\bottomrule
\end{tabular}
\end{center}
\begin{tablenotes}
    \footnotesize
    \item The better one is highlighted. All comparisons are statistically significant ($p<$.05 on Wilcoxon signed-rank test) and with large effect sizes (on $A_{12}$).
\end{tablenotes}
\end{table}

As can be seen clearly from Table~\ref{tb:mo-rq1}, over 30 repeated runs, NSGA-II is slightly better on the three objectives case while IBEA is better on the five objective case, in which cases the volume differences are relatively high. This is mainly due to the fact that the Pareto-dominance guided search in NSGA-II liked algorithms cannot scale well on more than three objectives, which has been theoretically analyzed in a large number of existing work, e.g., \cite{Li2015}. However, if the experimental comparison has not been guided by such a theoretical understanding, it is likely that only the three-objective case is compared. This could result in a misleading conclusion that \textit{``the experiments show the NSGA-II is better than IBEA on the SAS problem considered and thus it is chosen to derive subsequent study"} without taking the number of objectives into account, hence giving a wrong implication that the same can be applied to whatever number of objectives under the SAS problem. In this case, the addition of theoretical justification could easily motivate the need to validate cases beyond three objectives, or explicitly state that the conclusion may not be applicable to other cases with a different number of objectives to be considered. This is the possible issue we observed from some studies, such as~\cite{DBLP:conf/icsa/CalinescuCGKP17,DBLP:journals/ase/GerasimouCT18,DBLP:conf/saso/FredericksGK019}, where NSGA-II has been used with experimental justification only, and the conclusion implies that the approach (based on NSGA-II) can work equally well on SASs with more than three objectives. Another likely issue of $L_3$ is that even though more than three objectives have been considered, the lack of theoretical justification can lead to a comparison between similar and equally unfitted algorithms, such as the FastPGA and NSGA-II compared in~\cite{DBLP:journals/taas/Garcia-GalanPTC16}. Despite the fact that they both suffer from the issue of Pareto dominance when the number of objectives is greater than three, 
one may draw a conclusion that NSGA-II works better and thus should be used for SAS under such case, 
which is even more misleading. 
All the above have shown that a lack of justification for choosing the search algorithm can raise serious consequences to the field.

In fact, to justify the choice of the algorithm in SBSE for SASs, theoretical justification can often easily guide the design of experimental justification, but the opposed is difficult unless extensive empirical studies have been conducted. From the above, what we have shown is that choosing an algorithm based only on theoretical understanding may still cause issues, as some factors could be misconsidered. This is, nevertheless, less serious than choosing one based on experimental comparisons without theoretical justification, in which case misleading conclusions may be drawn easily. These are the reasons why they are ranked as $L_2$ and $L_3$, respectively. Clearly, $L_4$ is the worst case that should be avoided and its consequence could be dreadful. For example, a very recent work in SBSE for SASs~\cite{DBLP:journals/corr/abs-2004-11793}, which is ranked as $L_4$, has wrongly adapted NSGA-II to optimize a single-objective problem for SAS.

We would like to stress that our goal here is not to show an algorithm can be better than another in general, but to demonstrate the likely issues when choosing a search algorithm without proper justification in SBSE for SASs.

\subsubsection{Suggestion and Opportunity} 

Our suggestion is, therefore, simple:

\begin{tcolorbox}[breakable,left=5pt,right=5pt,top=5pt,bottom=5pt] 
\textbf{Suggestion 1:} If permitted, theoretically justifying the algorithm choice supported by experimental comparison ($L_1$), or at least, theoretical justification is a must ($L_2$). In all cases, avoiding the omission of justification or making choice solely according to analogy such as ``this algorithm is widely used" ($L_4$).
\end{tcolorbox}

The intrinsic reason behind the disappointment from Section~\ref{sec:rq1-di} is the lack of guidelines on choosing search algorithms for SASs. This raises a promising research opportunity:

\begin{tcolorbox}[breakable,left=5pt,right=5pt,top=5pt,bottom=5pt] 
\textbf{Opportunity 1:} Generic guidance on justifiably choosing search algorithms according to the requirements of the particular SAS problem studied. 
\end{tcolorbox}

Indeed, 
every search algorithm has its own merit,
which makes them well-suited to a particular class of SAS problems. For example, HC starts from a random initial point in the search space and 
iteratively finds out the best neighbor of the current solution, which could fit well with the planning problem for most service-based SASs as it is straightforward to design the `neighbor' based on the service providers. This feature helps converge fast if the search is in the ``right" direction, but may also cause it to get trap in local optimum easily. In contrast, the population-based search algorithm with diversity preservation, such as the GA, can help jump out of local optimum. It could be desirable for SAS with a large search space and complex types of variation points, e.g., planning for \textsc{RUBiS} and \textsc{SQLite}. However, such relatively random exploration can cause its slow convergence when there is a strict requirement for planning time.

The same applies to the Pareto search case. For example, 
the algorithms which compare solutions by Pareto dominance and density, such as NSGA-II~\cite{DBLP:journals/tec/DebAPM02} and SPEA2~\cite{Zitzler2002}, 
typically do not work well on many-objective problems where the number of objectives is larger than three~\cite{Purshouse2007}, which is not uncommon for SASs~\cite{DBLP:conf/icse/Garcia-GalanPTC14,DBLP:journals/taas/Garcia-GalanPTC16}. 
The decomposition-based algorithms (e.g., MOEA/D~\cite{Zhang2007} and its variants MOEA/D-STM~\cite{DBLP:journals/tec/LiZKLW14} and NSGA-III~\cite{Deb2014}) scale up well in terms of objective dimensionality, 
but may struggle on problems with an irregular Pareto front shape (e.g., degenerate, disconnect or highly-nonlinear)~\cite{Ishibuchi2017} --- a typical case between the throughput and cost objectives on SASs~\cite{DBLP:conf/gecco/0001LY18,DBLP:journals/infsof/ChenLY19}.
The indicator-based algorithms (e.g., IBEA~\cite{Zitzler2004}) are often insensitive to the Pareto front shape
but may suffer from the dominance resistance solutions (i.e. solutions with an extremely poor value on at least one of the objectives and (near) optimal values on the others~\cite{Ikeda2001}), thereby with their solutions concentrating in the boundaries of the Pareto front~\cite{Li2018}. This may produce some undesired effects when penalty terms exist in the requirements of SAS.

The linkage between the characteristics of search algorithms and the requirements of SAS problems is a unique challenge of this research opportunity. Such a linkage lies in the heart of the guidance to enable well-justified choices of search algorithms, supported with both theoretical and experimental justifications.
To achieve that, 
there are several research questions desirable for us to address.

\begin{itemize}

\item What kinds/form of requirements from the SAS problem can be important to the search process, such as the planning time, diversity of solutions, and self-adaptation purpose.

\item What characteristics of a search algorithm can better fit with such requirements, which can greatly inform the algorithm choice.

\item Or further, to determine in what context a further specialized search algorithm is a must.  

\end{itemize}

   \begin{figure}[!t]
  \centering
\begin{tikzpicture}
\begin{axis}[
  ybar=0.05cm,
  ymin=0,
  ymax=9,
     width=13cm,
    height=5cm,
    bar width=5pt,
    enlargelimits=0,
     enlarge x limits=0.05,
    legend style={at={(0.2,0.96)},
      anchor=north,legend columns=2},
    ylabel={\# primary studies},
     nodes near coords,
     nodes near coords={\pgfmathfloatifflags{\pgfplotspointmeta}{0}{}{\pgfmathprintnumber{\pgfplotspointmeta}}},
    symbolic x coords={2009,2010,2011,2012,2013,2014,2015,2016,2017,2018,2019},
     x tick label style={rotate=45,anchor=east},
    xtick=data
    ]

\addplot[fill=blue!50,draw=black] coordinates {(2009,2) (2010,3) (2011,4) (2012,2) (2013,2) (2014,6) (2015,2) (2016,5) (2017,4) (2018,6) (2019,8)};

\addplot[fill=yellow!50,draw=black] coordinates {(2009,0) (2010,0) (2011,1) (2012,1) (2013,3) (2014,1) (2015,2) (2016,2) (2017,3) (2018,1) (2019,2)};

\addplot[fill=red!50,draw=black] coordinates {(2009,0) (2010,0) (2011,0) (2012,0) (2013,0) (2014,1) (2015,3) (2016,2) (2017,2) (2018,6) (2019,4)};

\addplot[fill=white,draw=black] coordinates {(2009,0) (2010,0) (2011,0) (2012,2) (2013,0) (2014,0) (2015,1) (2016,0) (2017,0) (2018,0) (2019,0)};

\legend{Weights, Single, Pareto, Hierarchy}

\end{axis}

\end{tikzpicture}
   \caption{Popularity evolution on the formulation of search in SBSE for SASs (Seven studies use more than one types of formulations).}
 \label{fig:mo-rel}
  \end{figure}
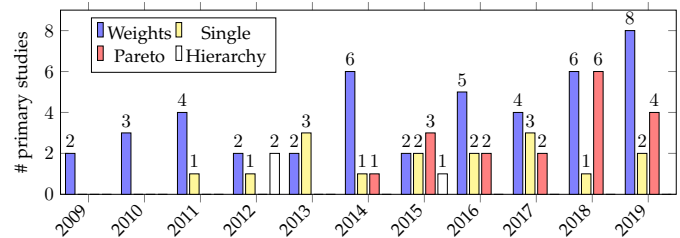

\subsection{RQ2: Objective Formulation in the Search for SASs}
\label{sec:rq2}

\subsubsection{Significance}

When engineering SBSE for SASs, another fundamental aspect is to determine the objective to be searched. As a result, understanding what, how, and why objectives are defined and handled during the search is crucial in SBSE for SASs, especially in the presence of multiple conflicting objectives. 

\subsubsection{Findings}
\label{sec:rq2-findings}

As illustrated in Figure~\ref{fig:mo-rel}, our reviews reveal that over the years, only a relatively small proportion of the studies consider the single-objective case. For the majority that take multiple objectives into account, weighted search, which combines all objectives via certain form of weighting strategy that effectively turns the problem into a single objective one\footnote{We found weighted sum and weighted product.} (a.k.a. utility-driven search), is the most predominant way to formulate the objectives in the search for SAS, such as~\cite{DBLP:conf/icac/RamirezKCM09,DBLP:conf/IEEEscc/MiWYZSY10,DBLP:conf/icse/FredericksDC14,DBLP:conf/saso/PodolskiyMKGP19,DBLP:journals/tse/WangHYY18}. Pareto search is ranked the second, e.g.,~\cite{DBLP:journals/tosem/ChenLBY18,DBLP:journals/jss/CalinescuCGKP18,DBLP:journals/tsc/ChenB17,DBLP:journals/taas/Garcia-GalanPTC16}, while hierarchical search, i.e., explicitly search one objective before another which is another way of objective aggregation, forms the minority~\cite{DBLP:journals/tse/RosaRLHS13,DBLP:conf/re/PengCYZ10,DBLP:conf/sigsoft/FilieriHM15}. There is indeed a tendency that the number of studies considers Pareto search to increase gradually since 2014. This, however, remains much less common compared with its weighted counterparts, especially in 2019. We have also found seven studies, e.g.,~\cite{DBLP:conf/wosp/0001BWY18,DBLP:conf/icse/KinneerCWGG18,DBLP:journals/ase/GerasimouCT18} exploit multiple types of objective formulation in the search, mostly due to they are used on different problems/aspects or contexts of the SAS. 

\begin{figure}[!t]
  \centering
  \begin{subfigure}[t]{0.525\columnwidth}
\begin{tikzpicture}
\begin{axis}[
    ybar stacked,
	bar width=10pt,
	nodes near coords, 
	nodes near coords={\pgfmathfloatifflags{\pgfplotspointmeta}{0}{}{\pgfmathprintnumber{\pgfplotspointmeta}}},
    enlargelimits=0,
    enlarge x limits=0.05,
    legend style={at={(0.65,0.99),font=\large},
      anchor=north,legend columns=1},
    ylabel={\# primary studies},
    xtick={2009,2010,...,2019},
    y label style={at={(-0.08,0.5)},font=\large},
    symbolic x coords={2009, 2010, 2011, 2012, 
		2013, 2014, 2015, 2016, 2017, 2018, 2019},
       x tick label style={rotate=45,anchor=east,font=\large},
    y tick label style={font=\large}
    ]
\addplot[ybar, fill=white, draw=black] plot coordinates {(2009,1) (2010,0) 
  (2011,1) (2012,0) (2013,0) (2014,3) (2015,1) (2016,0) (2017,1) (2018,0) (2019,0)};
  
  \addplot[ybar, fill=blue!35, draw=black] plot coordinates {(2009,1) (2010,3) 
  (2011,3) (2012,2) (2013,2) (2014,3) (2015,1) (2016,5) (2017,3) (2018,6) (2019,8)};

\legend{With reasons,Unknown}

\end{axis}
\end{tikzpicture}
    \subcaption{Weighted search}
 \label{fig:alg-s-evo}
    \end{subfigure}
    \hspace{-0.2cm}
      \begin{subfigure}[t]{0.475\columnwidth}
\begin{tikzpicture}
\begin{axis}[
    ybar stacked,
	bar width=10pt,
	ymax=7,
	nodes near coords, 
	nodes near coords={\pgfmathfloatifflags{\pgfplotspointmeta}{0}{}{\pgfmathprintnumber{\pgfplotspointmeta}}},
    enlargelimits=0,
     enlarge x limits=0.5,
    legend style={at={(0.35,0.99),font=\large},
      anchor=north,legend columns=1},
    symbolic x coords={2014, 2015, 2016, 2017, 2018, 2019},
    xtick=data,
    x tick label style={rotate=45,anchor=east,font=\large},
    y tick label style={font=\large}
    ]
\addplot[ybar, fill=white, draw=black] plot coordinates {(2014,0) (2015,1) (2016,2) (2017,1) (2018,5) (2019,3)};
  
  \addplot[ybar, fill=blue!35, draw=black] plot coordinates {(2014,1) (2015,2) (2016,0) (2017,1) (2018,1) (2019,1)};

\legend{With reasons,Unknown}

\end{axis}
\end{tikzpicture}
    \subcaption{Pareto search}
 \label{fig:alg-m-evo}
    \end{subfigure}
    \caption{Evolution of whether reasons have been provided when using weighted search and Pareto search for SAS.}
    \label{fig:alg-s-m-evo}
  \end{figure}
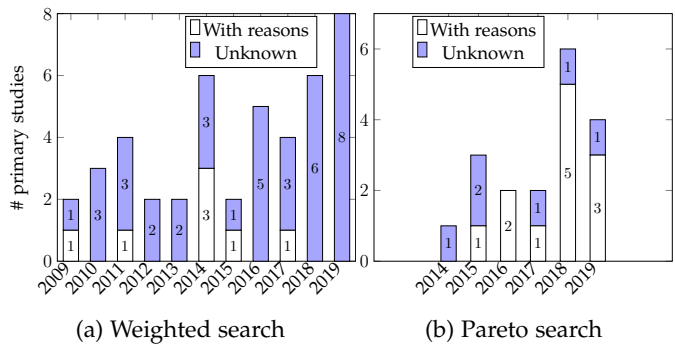

Since considering multiple objectives are more pragmatic, in which weighted search and Pareto search are the two most popular (yet alternative) ways of formulating SBSE for SASs, in Figure~\ref{fig:alg-s-m-evo}, we summarized the reason behind their choice of the two formulations. Clearly, for the weighted case, the majority of the choice gives no clear reasons, e.g.,~\cite{DBLP:conf/sigsoft/MaggioPFH17}. For those that do, the most common reason is that \textit{``the weights can flexibly allow one to specify the relative importance between objectives"}~\cite{DBLP:conf/sigsoft/MaggioPFH17}. Pareto search, in contrast, often provide clear reasons. For example, Gerasimou et al.~\cite{DBLP:journals/ase/GerasimouCT18} explain that Pareto search is chosen because it reveals richer information about the trade-offs between multiple QoS requirement, leading to better-informed decision making for SASs. Table~\ref{tb:mo-so} shows the treatments and assumptions applied to these two formulations of search under multi-objectivity. For weighted search, the weights are often left to the engineers to provide in a prior (28 studies), such as~\cite{DBLP:journals/fgcs/BarakatML18}; or equal weights are assumed by default to reflect equal importance (16 studies) which is believed to achieve a balanced outcome~\cite{DBLP:conf/sigsoft/EsfahaniKM11}, such as~\cite{DBLP:conf/IEEEscc/MiWYZSY10}. For Pareto search, the final trade-off solution is most commonly left to the engineers (e.g.,~\cite{DBLP:conf/kbse/GerasimouTC15,DBLP:conf/saso/FredericksGK019}) while three studies~\cite{DBLP:journals/tsc/ChenB17,DBLP:conf/wosp/0001BWY18,DBLP:journals/tosem/ChenLBY18} automatically select the knee solution --- the solution that achieves a balanced result without explicit weights. Interestingly, some studies, such as Gal{\'{a}}n et al. ~\cite{DBLP:journals/taas/Garcia-GalanPTC16}, apply an additional weighted function to select the final one from the solution set produced by Pareto search, but the reason of which has not been discussed.

As for the actual search objectives, Table~\ref{tb:objectives} shows a summary of the most common ones. Overall, latency and cost are the most overwhelmingly targeted objectives individually whilst in terms of objective combination on SASs, latency and cost are also the most predominately targeted case. It is worth noting that some combinations are clearly conflicting, such as latency and cost; latency and power. Some others tend to be more harmonic, such as latency and throughput. Constraints are also sometime considered, in which the most common ones are threshold (e.g., threshold of latency requirements)~\cite{DBLP:journals/jss/BashariBD18} and dependency between variation points~\cite{DBLP:conf/icse/PascualPF13,DBLP:journals/tosem/ChenLBY18}. For example, the \texttt{cache\_mode} cannot be changed until \texttt{cache} option has been enabled. We would like to stress that the constraints may also be considered as an objective depending on the search algorithms used, for example, Chen et al. ~\cite{DBLP:journals/tosem/ChenLBY18} consider dependency as a constraint but Gal{\'{a}}n et al. ~\cite{DBLP:journals/taas/Garcia-GalanPTC16} treat such as an objective. As a summary, our findings for \textbf{RQ2} include:

\begin{tcolorbox}[breakable,left=5pt,right=5pt,top=5pt,bottom=5pt] 
\textbf{Findings 4:} Multiple objectives case is much more common than single-objective assumption in SBSE for SASs, in which the weighted search is predominately used. \\
\textbf{Findings 5:} The formulation behind weighted search often provide no reasons while that for Pareto search is usually discussed in details.\\ 
\textbf{Findings 6:} A vast set of objectives and their combination have been targeted; a constraint in one study may also be used as objectives in some other studies.
\end{tcolorbox}

 \begin{table}[t!]
\centering
\caption{Treatments in Pareto-based and weighted search on SBSE for SAS.}
\label{tb:mo-so}

\centering

\begin{tabular}{cc}\toprule

\makecell{Weighted\hfill}&\makecell{\textbf{How to specify weights in SAS?}\\Given by engineers (28), Equal weights (16),\\Dyanamic update (1), Probability (2)}\\ 
\hline
\makecell{Pareto-based}&\makecell{\textbf{How to make trade-off for SAS?}\\Leave to engineers (12), Knee solution (3),\\Weighted choice (2), Preferred objective(s) (1)}\\

\bottomrule
\end{tabular}

\begin{tablenotes}
    \footnotesize
    \item Number in the bracket indicates how many studies are involved.
\end{tablenotes}
\end{table}
  
  \begin{table}[t!]
\caption{Top 5 objective and constraint on SBSE for SAS.}
\label{tb:objectives}
{
\centering

\begin{tabular}{ccc}\toprule

\textbf{SAS Objective}\hfill&\textbf{Combination}&\textbf{SAS Constraint}\\ 
\midrule

Latency$^{+}$ (25)&Latency and cost (16)&Threshold$^{\star}$ (11)\\
Cost (23)&Latency and reliability (9)&Dependency$^{\star}$ (7)\\
Power (12)&Latency and throughput (7)&Resources$^{\star}$ (3)\\
Utility (10)&Latency and power (5)&Reliability$^{\star}$ (1)\\
Reliability$^{\star}$ (9)&Cost and throughput (3)&Ordering (1)\\

\bottomrule
\end{tabular}
}
\begin{tablenotes}
    \footnotesize
    \item Number in the bracket indicates how many studies are involved.
    \item $^{+}$ can refer to response time or performance.
    \item $^{\star}$ means it has also been used as objective/constraint in other studies.
\end{tablenotes}
\end{table}

\subsubsection{Disappointments}
\label{sec:rq2-di}

When dealing with multiple objectives in SAS, it is disappointed to find that the weighted search is much more commonly used than its Pareto counterpart in SBSE for SASs, albeit the latter is regarded as the better option in offering an understanding of the search problem at the SBSE community~\cite{DBLP:journals/csur/HarmanMZ12}. We also found that such a trend is prevalent across the years with little changes, as shown from Figure~\ref{fig:mo-rel}.

In essence, the aggregation of objectives implies that certain preferences between the objectives are available and they can be precisely quantified using weights. We showed that the majority of them have either assumed the weights can be provided by the engineers or they are equally weighted by default. However, as widely recognized from the literature~\cite{DBLP:journals/csur/HarmanMZ12,DBLP:journals/tmc/PaolaFGRD17,DBLP:journals/tosem/ChenLBY18}, it is not uncommon that a clear and precise quantification of the weights is very difficult, if not impossible, especially given the complexity of SAS. This is what should have been justified when assuming the weighted search for SASs, which unfortunately we have failed to see in most studies.

Of course, if the search space is so small and the evaluation is rather cheap on a given SAS problem, then it does not really matter which formulation to use as all the solutions can be identified and searched easily. However, this needs to be discussed explicitly to justify that the choice of objective formulation in the search has no impact. A further evidence to this disappointment is that majority of the studies that adapt weighted search has no discussion on the reasons behind --- only 7 out of 43 studies have explained reasons, such as the weights can be explicitly given because of the special characteristics of the SAS problem/subject domain considered~\cite{DBLP:conf/icac/RamirezKCM09}. This is by contrast to the 12 cases (out of 18) supported with reasons when formulating the search in a Pareto manner. In fact, a considerable amount of studies~\cite{DBLP:journals/tmc/PaolaFGRD17,DBLP:journals/tosem/ChenLBY18,DBLP:conf/gecco/0001LY18} have provided the reasons of using Pareto search by clearly comparing with the weighted search, but we found none of the similar cases when the weighted search is chosen. The wide adoption of weighted search without justification, especially given its clear limitations, can cause threats to the validity and applicability of the work in SBSE for SASs. Our disappointment is, therefore:

\begin{tcolorbox}[breakable,left=5pt,right=5pt,top=5pt,bottom=5pt] 
\textbf{Disappointment 2:} Unjustified and limited formulation on the multi-objective search for SASs. 
\end{tcolorbox}

\subsubsection{Justification on the Likely Issues} 

A clear advantage of using Pareto search on multiple SAS objectives is the fact that it does not require weights specification. In addition to this, the unique search setting to approximate the Pareto front may make it possible to discover some irregular search regions that would be otherwise difficult to be found with the weighted search.

To justify why it could be important to consider Pareto search as opposed to the current trend of SBSE for SASs where the weighted search is predominately used, we experimentally compare how NSGA-II and GA perform when optimizing the SASs, as the representative of Pareto and weighted search under equal weights (with normalization), respectively. The reason why we have chosen these two is due to their algorithmic similarity, i.e., we can then focus on the formulation of the search they rely upon. They all use the same parameter settings, e.g., population size and evaluations, details of which can be found in the supplementary. We run them on three SASs, \textsc{LLVM}, \textsc{Trimesh} and a service-based system (\textsc{SBS}), under the scenario of design-time profiling. Three subject SASs from distinct domains achieve better coverage than 87\% of the current studies (as what will be shown in Section~\ref{sec:rq5}) and have been used in prior work~\cite{DBLP:journals/corr/abs-1801-02175,DBLP:journals/tosem/ChenLBY18}. Details of the SASs can also be found in the supplementary. In particular, these SASs are chosen because they involve two objectives to be tuned and their objective spaces are diverse, with different shapes and densities of the trade-off surface. A total of 100 runs have been conducted.

\begin{figure}[t!]
\centering
\includegraphics[width=\columnwidth]{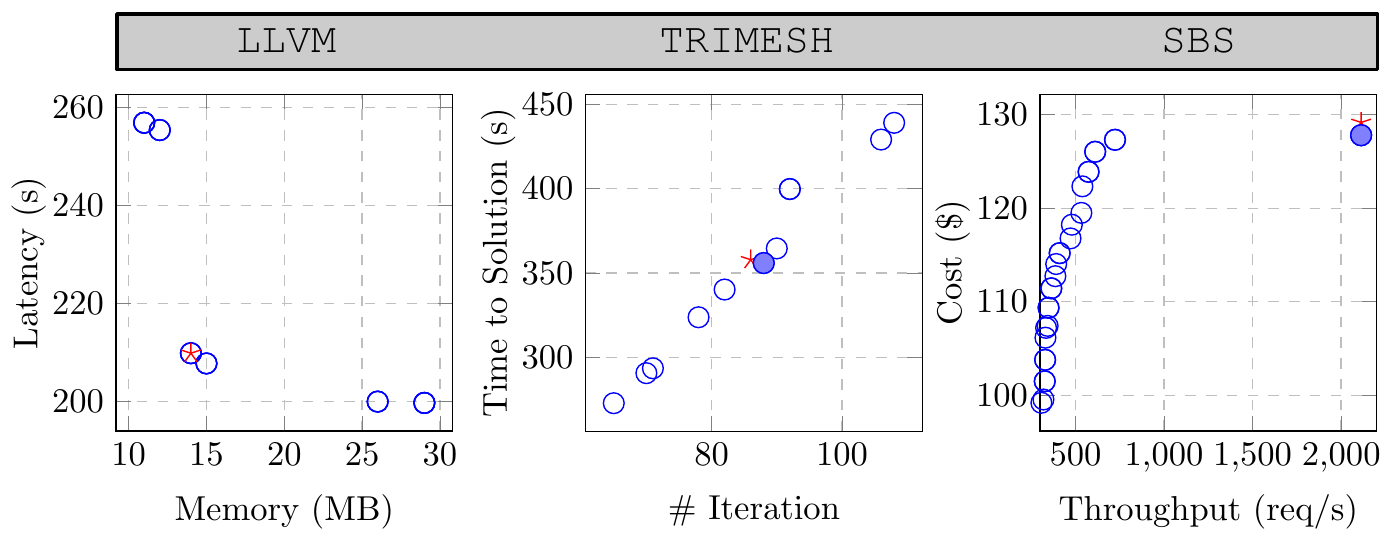}
\caption{Examples of the common results between Pareto (NSGA-II denoted as \copy\aMark) and equally weighted search [0.5,0.5] with normalization (GA denoted as \copy\bMark) for three SASs under a workload, job, or service quality/availability change (\# iterations and throughput are to be maximized while others are to be minimized).}
\label{fig:w-vs-p}
\end{figure}

The resulted objective space of one example, which is common across all the runs, has been shown in Figure~\ref{fig:w-vs-p}. From this, we can obtain the following observations:

\begin{enumerate}
    \item All three subject SASs reveals that the weighted search, albeit being proceeded by a population-based algorithm like GA, would converge to one point. Pareto search would approximate the whole Pareto front by contrast.
    \item The \textsc{Trimesh} and \textsc{SBS} cases reveal that the sole solution produced by weighted search can be dominated by the solutions found by Pareto search. This explains why we found that certain studies~\cite{DBLP:conf/icse/Garcia-GalanPTC14,DBLP:journals/taas/Garcia-GalanPTC16} apply both searches for the same SAS problem and context: because the weighted search may be over-constrained by the given weights, hence struggling to find some good solutions in the first place.
    \item \textsc{SBS} implies that, albeit being assumed in 16 studies, an equal weight may not lead to a balanced outcome depending on the shape of Pareto front. In fact, it could largely bias towards a certain objective, which contradicts with the most common reason for using equally weighted search for SAS under weights/utility theory~\cite{DBLP:conf/sigsoft/EsfahaniKM11}. 
\end{enumerate}

The above has, therefore, justified that the overwhelming adoption of the weighted search on SAS, especially when there is a lack of justification, is problematic to the field.

Indeed, one challenge of Pareto multi-objective search is how to select a single solution from the produced non-dominated trade-off set. When no preference information is available for the design-time problem, it is ideal to provide the engineers with all non-dominated solutions found to keep them informed in the decision-making process. When certain preferences exist (or at runtime), the selection of the single solution can be tailored with such or according to some assumptions of the preferences in case of runtime problem. Yet, unlike the case of weighted search, such preferences do not require explicit quantification. As we have shown in Section~\ref{sec:rq2-findings}, this can be automatically completed by selecting a solution from certain regions, e.g., the knee point selection.

\subsubsection{Suggestion and Opportunity}

According to the above, our suggestions to overcome the disappointment from Section~\ref{sec:rq2-di} in the presence of multiple SAS objectives is apparently:

\begin{tcolorbox}[breakable,left=5pt,right=5pt,top=5pt,bottom=5pt] 
\textbf{Suggestions 2:} When dealing with multiple objectives on SASs, always consider Pareto search as a possible alternative, regardless whether weights can be explicitly set. 
\end{tcolorbox}

A particular opportunity which is now under-explored in SBSE for SASs is:

\begin{tcolorbox}[breakable,left=5pt,right=5pt,top=5pt,bottom=5pt] 
\textbf{Opportunity 2:} Pareto many-objective search for SASs. 
\end{tcolorbox}

Conventionally, Pareto many-objective search targets the case when the number of objectives is greater than three. Similar to the classic Pareto multi-objective search, such a paradigm is also free from the tedious weight specification but aims to explicitly overcome the limitation introduced by Pareto-dominance guided search. This fits well with the requirements of SAS, in which case the current treatment relies on weighted search.

Unlike some other SBSE problems, the unique property in a SAS problem with different objectives is that the relations between these objectives may not necessarily be conflicting, or only partially be conflicting depending on the environment. For example, as we have shown, the widely used pair of objectives, such as latency and throughput, tends to be more harmonic. When the objective number is small, this may not be an issue as the dimensionality may not cause too much challenge to the selection pressure. However, for Pareto many-objective search, such a unique property of SAS problems could be better exploited and specialized in the algorithm.

There are already readily available Pareto many-objective search algorithms~\cite{DBLP:conf/cec/IshibuchiTN08}, but it is yet clear how they can be specialized for SASs to better meet the requirements of a SAS problem. To this end, the key challenges of this research opportunity are, therefore:

\begin{itemize}

    \item Which SAS problem, MAPE-K phases, and self-adaptation propose can suit well with the pros and/or cons of Pareto many-objective search.
    
    \item How to consider SAS problem-specific objective relation (conflicting or harmonic) in the search and solution selection, especially when there is a high dimensional objective space.

    
    
    \item How to mitigate the rapidly increasing search cost (i.e., space and time) to fit the timeliness requirements of certain SAS problems at runtime.
\end{itemize}

\subsection{RQ3: Evaluating the Pareto Search for SASs}
\label{sec:rq5}

\subsubsection{Significance}

When optimizing SAS that involves only a single or aggregated objective, the quality of the SBSE approach can be simply evaluated by using that objective value or the given weight vector. However, in the case that Pareto search is involved, selecting appropriate quality indicator(s), which assess the quality of a solution set produced, becomes a critical yet challenging task since different solutions may be incomparable on the basis of Pareto dominance. Therefore, it is of great significance to understand what types of methods/indicators are currently used to serve such purpose and the reasons behind it.

Recall from Table~\ref{tb:objectives}, it is often the case that the objectives to be optimized by a search algorithm are directly related to the ultimate quality concerns of the SAS that the overall approach seeks to improve. Therefore, the evaluation of the quality concerns in SAS is equivalent to the evaluation of the search algorithm, which is an integral part of the SBSE approach. This is particularly true for the studies that adopt Pareto search, wherein the exact quality concerns of the SAS are searched/optimized directly by the algorithm. 

\subsubsection{Findings}

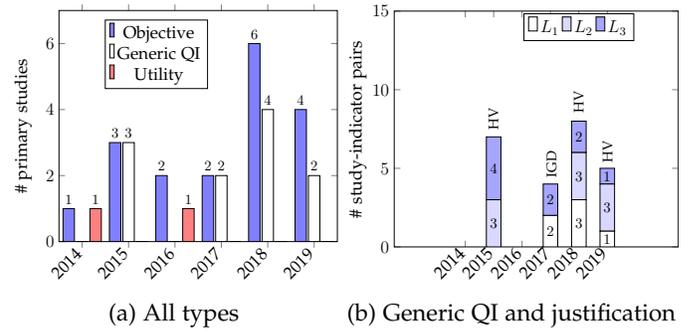
\begin{figure}[!t]
  \centering
  \begin{subfigure}[t]{0.49\columnwidth}
\begin{tikzpicture}
\begin{axis}[
  ybar=0.05cm,
  ymin=0,
  ymax=7,
    bar width=8pt,
    enlargelimits=0,
     enlarge x limits=0.1,
    legend style={at={(0.35,0.96),font=\large},
      anchor=north,legend columns=1},
         y label style={at={(-0.08,0.5)},font=\large},
    ylabel={\# primary studies},
     nodes near coords,
     nodes near coords={\pgfmathfloatifflags{\pgfplotspointmeta}{0}{}{\pgfmathprintnumber{\pgfplotspointmeta}}},
    symbolic x coords={2014,2015,2016,2017,2018,2019},
       x tick label style={rotate=45,anchor=east,font=\large},
    y tick label style={font=\large},
    xtick=data
    ]

\addplot[fill=blue!50,draw=black] coordinates {(2014,1) (2015,3) (2016,2) (2017,2) (2018,6) (2019,4)};

\addplot[fill=white,draw=black] coordinates {(2014,0) (2015,3) (2016,0) (2017,2) (2018,4) (2019,2)};

\addplot[fill=red!50,draw=black] coordinates {(2014,1) (2015,0) (2016,1) (2017,0) (2018,0) (2019,0)};

\legend{Objective, Generic QI, Utility}

\end{axis}

\end{tikzpicture}
    \subcaption{All types}
  \label{fig:qi-count}
    \end{subfigure}
    \hspace{-0.1cm}
      \begin{subfigure}[t]{0.5\columnwidth}
\begin{tikzpicture}
\begin{axis}[
    ybar stacked,
	bar width=10pt,
	ymax=15,
	nodes near coords, 
	nodes near coords={\pgfmathfloatifflags{\pgfplotspointmeta}{0}{}{\pgfmathprintnumber{\pgfplotspointmeta}}},
    enlargelimits=0,
     enlarge x limits=0.5,
    legend style={at={(0.65,0.99),font=\large},
      anchor=north,legend columns=-1},
        ylabel={\# study-indicator pairs},
    y label style={at={(-0.08,0.5)},font=\large},
    symbolic x coords={2014, 2015, 2016, 2017, 2018, 2019},
    xtick=data,
    x tick label style={rotate=45,anchor=east,font=\large},
    y tick label style={font=\large}
    ]
\addplot[ybar, fill=white, draw=black] plot coordinates {(2014,0) (2015,0) (2016,0) (2017,2) (2018,3) (2019,1)};
  
  \addplot[ybar, fill=blue!15, draw=black] plot coordinates {(2014,0) (2015,3) (2016,0) (2017,0) (2018,3) (2019,3)};
  
    \addplot[ybar, fill=blue!35, draw=black] plot coordinates {(2014,0) (2015,4) (2016,0) (2017,2) (2018,2) (2019,1)};

\legend{$L_1$, $L_2$, $L_3$, $L_4$}

\end{axis}
\coordinate (A) at (1.7,1.2);
\coordinate (B) at (2.4,3.1);
\coordinate (C) at (3.15,2.7);
\coordinate (D) at (3.8,2);
\coordinate (E) at (4.45,3.5);
\coordinate (F) at (5.15,2.3);

\node[rotate=90] at (B) {HV};
\node[rotate=90] at (D) {IGD};
\node[rotate=90] at (E) {HV};
\node[rotate=90] at (F) {HV};
\end{tikzpicture}
    \subcaption{Generic QI and justification}
  \label{fig:qi-all}
    \end{subfigure}
      \caption{Evolution of evaluation methods to assess Pareto SBSE for SAS (All studies use individual objective values; 11 use more than one type).}
  \end{figure}

For the 18 studies that consider Pareto search for SASs, 
Figure~\ref{fig:qi-count} depicts the types of quality evaluation methods and their popularity to assess solution sets over years. As can be seen, examining directly on each objective is the most common way (e.g., reporting the mean or plotting the results), following by generic quality indicators that were designed specifically to evaluate solution sets, such as HV~\cite{Zitzler1998}, GD~\cite{Veldhuizen1998} and IGD~\cite{Coello2004}. 
Two studies~\cite{DBLP:conf/icse/Garcia-GalanPTC14,DBLP:journals/taas/Garcia-GalanPTC16} leverage on a given weight vector to evaluate the solutions produced by the Pareto search. 
Such a trend remains unchanged as the field evolves. We note that 11 out of the 18 studies, e.g.,~\cite{DBLP:journals/tosem/ChenLBY18,DBLP:journals/jss/CalinescuCGKP18,DBLP:journals/infsof/ChenLY19}, 
have used different types of methods to assess Pareto SBSE for SASs, due primarily to their complementary nature. 
For example, plotting all the objective values can be a good addition to the results from generic quality indicators~\cite{DBLP:journals/csur/LiY19}.

    \begin{figure}[!t]
  \centering
\begin{tikzpicture}
\begin{axis}[
 xbar stacked,
    width=7cm,
    height=5cm,
    xmin=0,
    bar width=0.25cm,
    	nodes near coords, 
	nodes near coords={\pgfmathfloatifflags{\pgfplotspointmeta}{0}{}{\pgfmathprintnumber{\pgfplotspointmeta}}},
  enlargelimits=0,
 enlarge y limits=0.05,
    legend style={at={(0.85,0.65)},
      anchor=north,legend columns=1},
    xlabel={\# primary studies},
    symbolic y coords={no,ED,GS,SP,C,CI,GD,epsilon,IGD,HV},
     yticklabels={Hypervolume \textbf{(HV)},Inverted Generational Distance \textbf{(IGD)},$\epsilon$-Indicator \textbf{($\epsilon$)},Generational Distance \textbf{(GD)},Contribution Indicator \textbf{(CI)},$\mathcal{C}$ Metric \textbf{($\mathcal{C}$)},Spacing \textbf{(SP)},General Spread \textbf{(GS)},Euclidean Distance \textbf{(ED)},No generic quality indicator used  
     },
         ytick=data
    ]
\addplot[xbar, fill=white, draw=black] coordinates {(2,HV) (2,IGD) (2,epsilon) (0,GD) (0,CI) (0,C) (0,SP) (0,GS) (0,ED) (0,no)};

\addplot[xbar, fill=blue!15, draw=black] coordinates {(2,HV) (2,IGD) (2,epsilon) (1,GD) (0,CI) (0,C) (1,SP) (1,GS) (0,ED) (0,no)};

\addplot[xbar, fill=blue!35, draw=black] coordinates {(4,HV) (1,IGD) (0,epsilon) (1,GD) (1,CI) (1,C) (0,SP) (0,GS) (1,ED) (7,no)};

\legend{$L_1$, $L_2$, $L_3$, $L_4$}
\end{axis}
\end{tikzpicture}
   \caption{The generic quality indicators (including cases when none of them are used) and their levels of justification in SBSE for SASs over years (Most studies use more than one indicator).}
 \label{fig:qi-top}
  \end{figure}
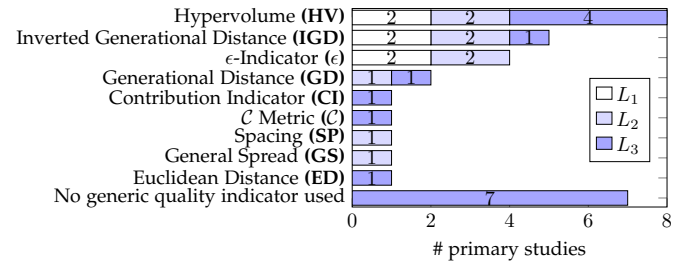

Unlike the other types, there are more than hundreds of generic quality indicators and thus their selection is also a challenge~\cite{DBLP:journals/csur/LiY19}. Figure~\ref{fig:qi-all} shows the levels of justification when choosing each generic quality indicator in SBSE for SASs, in which HV tends to be the most popular indicator used. We note that over the years, $L_3$ (e.g.,~\cite{DBLP:journals/jss/PascualLPFE15}) or $L_2$ (e.g.,~\cite{DBLP:journals/ase/GerasimouCT18}) are much more common than the $L_1$ cases~\cite{DBLP:conf/icsa/CalinescuCGKP17,DBLP:journals/infsof/ChenLY19} when justifying the choice. Indeed, we have found that most of the studies have used more than one indicator, but not all of them have a justification. The most common reason is that certain indicators only cover part of the quality aspect~\cite{DBLP:journals/csur/LiY19}, or there is a specific requirement according to the preferences of the SAS problem~\cite{DBLP:conf/icsa/CalinescuCGKP17}.

To ensure that such a trend is not biased by a particular indicator, Figure~\ref{fig:qi-top} plots all the generic quality indicators used and their levels of justification, together with the reasons for cases when no indicator is used at all. We note that while HV, IGD and $\epsilon$-indicator are much more popular than the others, the overall trend of justification levels remains the same as Figure~\ref{fig:qi-all}. In particular, seven studies (out of 18), such as~\cite{DBLP:conf/icse/KinneerCWGG18,DBLP:journals/taas/Garcia-GalanPTC16}, have not used any generic quality indicator, for which no justification has been provided. In summary, our findings on evaluating Pareto search for SASs under \textbf{RQ3} are:

\begin{tcolorbox}[breakable,left=5pt,right=5pt,top=5pt,bottom=5pt] 
\textbf{Findings 7:} Directly assessing on each objective remains the most common evaluation methods in Pareto search for SASs. \\
\textbf{Findings 8:} A considerable amount of studies have used none of the generic quality indicators. 
However, no justification is provided under these cases. \\
\textbf{Findings 9:} The choice of indicators often lies at the justification level $L_3$ or $L_2$.
\end{tcolorbox}

\subsubsection{Disappointments}
\label{sec:rq3-di}

Our disappointment lies in the fact that generic quality indicators, despite being rather successful in SBSE, remains far away from standard practice in SBSE for SASs~\cite{DBLP:conf/icse/Li0Y18}.

Indeed, plotting the objectives may seem to be a simple way for the assessment. Yet this only works well for the bi-objective case, and when the number of objectives reaches four or more 
(which is not uncommon for SASs), it is difficult to clearly illustrate the solution sets by scatter plot. More importantly, visual comparison cannot provide a quantitatively comparable result between the solution sets. Reporting the mean values of each objective may also seem straightforward, but they neither reflect the trade-off nor the overall quality of the solution set, 
leaving many aspects uncovered. 
Therefore, in conjunction with the above, generic quality indicators are promising to overcome these limitations~\cite{DBLP:conf/icse/Li0Y18,DBLP:journals/csur/LiY19}. 

Since the possible number of generic quality indicators is enormously high, a perhaps even more disappointing point is that the justification of the choice has been insufficient, i.e., only a small proportion can reach $L_1$. To give a good example of $L_1$, Calinescu et al.~\cite{DBLP:conf/icsa/CalinescuCGKP17} adopted IGD and $\epsilon$-indicator because they need to assess three quality aspects of a solution set (i.e., convergence, spread, and uniformity), which are all important for the SAS problem studied. In addition, there is a specific preference of robustness in the problem with respect to the quality aspects, and therefore these indicators are further tailored to fit such a need --- a typical example where the choice of indicators are driven by the quality aspects covered and their relations to the preference in the SAS problem. 

In fact, we found that most of the time the choices are solely driven by the analogy that other work has also used the same ones, which is a typical case of $L_3$. For example, Fredericks et al.~\cite{DBLP:conf/saso/FredericksGK019} used HV as the sole indicator simply because it is well-known in the field. This is of concern, as lack of justification (or even none at all) may result in misleading conclusions which we will discuss in Section~\ref{sec:rq3-justify}. Our findings have also confirmed that such a trend neither is due to bias on using a particular indicator nor tends to be changed over the years. The negligence of generic quality indicators, together with the limited justification of the choice, are severe threats to the conclusion validity. Overall, our disappointment can be summarized as:

\begin{tcolorbox}[breakable,left=5pt,right=5pt,top=5pt,bottom=5pt] 
\textbf{Disappointment 3:} Questionable choice of evaluation methods in Pareto search for SASs.
\end{tcolorbox}

\subsubsection{Justification of the Likely Issues}
\label{sec:rq3-justify}


The likely issue of our disappointment here can be caused by the fact that each indicator has, by design, its own assumption of preferences and the quality aspects covered, as shown in Table~\ref{tb:taxonomy}. This is precisely the reason why their choice cannot be done arbitrarily, as the one that fitted well in other situations may not be suitable for the SAS problem studied. For example, HV measures all the four quality aspects, which implies that one study uses HV only because all four quality aspects are of interest, such as the planning for SASs that seek to tune latency, reliability, and throughput under no preferences~\cite{DBLP:journals/tosem/ChenLBY18}. GS, which measures solely the diversity, could be an ideal indicator for SAS testing~\cite{DBLP:conf/icse/FredericksDC14} with two quality attributes of interest, such as latency and power, as the aim therein is to verify the behaviors of SAS by using a diverse set of test cases that covers different trade-off points between latency and power. As a result, when some of the chosen quality indicators do not agree with each other, it could be simply due to the fact that they assess different quality aspects of the solution set. This is the key reason why justification level at $L_3$ is problematic.

\begin{table}[t!]
	\setlength{\tabcolsep}{2mm}
		\caption{The generic quality indicators for assessing solution set in SBSE for SAS.}	
	        \label{tb:taxonomy}
		\begin{tabular}{ccccc}\toprule
		\textbf{Indicator}&\textbf{Convergence}&\textbf{Spread}&\textbf{Uniformity}&\textbf{Cardinality}\\
		\midrule     
		\textbf{GD}&$+$&&&\\\hline
		\textbf{ED}&$-$&&&\\\hline
		\textbf{$\epsilon$-indicator}&$+$&$+$&$+$&$-$\\\hline
		\textbf{GS}&&$-$&$+$&\\\hline
		\textbf{CI}&$-$&&&\\\hline
		\textbf{IGD}&$+$&$+$&$+$&$-$\\\hline
		\textbf{HV}&$+$&$+$&$+$&$-$\\\hline
		\textbf{SP}&&&$+$&\\\hline
		$\mathcal{C}$&$-$&&&$-$\\
			
			\bottomrule	
		\end{tabular}
		\begin{tablenotes}
      \footnotesize
      \item Diversity consists of spread (i.e., coverage) and uniformity.
			``$+$'' means that the indicator can well reflect a specific quality aspect and 
			``$-$'' means that the indicator can partially reflect a quality aspect.
    \end{tablenotes}	

\end{table} 

$L_2$ can still be insufficient because even for two indicators which are designed for assessing the same quality aspect of a solution set, they could work only on cases with very different preferences~\cite{DBLP:conf/icse/Li0Y18,DBLP:journals/csur/LiY19}, such as a region of interests, the priority of objectives or even subject to some vague constraints. For example, both HV and IGD are used to provide a comprehensive evaluation of a solution set in terms of convergence, spread, uniformity, and cardinality, but HV clearly prefers knee points of the Pareto front and IGD prefers uniformly-distributed solutions~\cite{DBLP:journals/csur/LiY19}. Therefore, a careful and justifiable selection and use of quality indicators to evaluate/compare solution sets have to be made in relation to the preferences of the SAS problems~\cite{DBLP:journals/csur/LiY19,DBLP:conf/icse/Li0Y18}. 

\begin{figure}[t!]
	\centering
	\begin{tikzpicture}
    \begin{axis}[
        grid = major,
        grid style={dashed},
        height=7cm,
        width=7cm,
        ymax=1.1,
        xmax=550,
        ylabel=Reliability (\%),
        xlabel=Latency (s),
           label style={font=\fontsize{12}{12}\selectfont},
                    tick label style={font=\fontsize{12}{12}\selectfont},
      legend style={font=\fontsize{12}{12}\selectfont,at={(0.9,0.2)}, anchor=east,legend columns=1}]


 \addplot[only marks,mark=o,blue,mark size=4pt] plot coordinates  {
 (0,0)
 (100,0.4)
  (200,0.7)
  (350,0.9)
  (500,1)
 };
 
 \addplot[only marks,mark=star,red,mark size=4pt] plot coordinates  {
 (200,0.2)
 (350,0.4)
 (400,0.6)
 (450,1)
 };

   \node[align=left,anchor=west,black] (source) at (axis cs:300,1){\fontsize{12}{12}\selectfont $\beta$};
   \node (destination) at (axis cs:430,1){};
    \draw[->][black,thick](source)--(destination);

\addlegendentry{A}
\addlegendentry{B}
    \end{axis}
    \end{tikzpicture}
	\caption{An example in SBSE for SASs where the generic quality indicators can be misleading. When searching for the minimal latency and the best reliability of SAS adaptation upon a workload change~\cite{DBLP:conf/kbse/GerasimouTC15,DBLP:journals/tse/WangHYY18,DBLP:journals/tsc/ChenB17}, two Pareto search algorithms produce two nondominated solutions sets, $A$ and $B$, respectively. $A$ is evaluated better than $B$ on all the nine commonly used quality indicators in SBSE for SASs (solutions being normalized before the evaluation):
	$GD(A)=0.02 < GD(B)=0.26, ED(A)=0.5 < ED(B)=0.89, \epsilon(A)=0.1 < \epsilon(B)=0.3, 
	GS(A)=0.15 < GS(B)=0.46, CI(A)=0.8 > CI(B) = 0.2,
	IGD(A)=0.02 < IGD(B)=0.27,
	HV(A)=0.77 > HV(B)=0.43, SP(A)=0.05 < SP(0.1),
	\mathcal{C}(A)=0.8 > \mathcal{C}(B)=0.25.$ 
	However, $B$ is, in fact, more preferred (specifically solution $\beta$) 
	when full reliability is more important than possible low latency.}
	\label{fig:qi-example}
\end{figure}
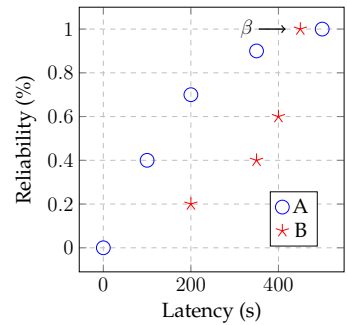

To justify the likely issues raised from $L_2$ level of indicator choice, we show an example SAS at Figure~\ref{fig:qi-example} using the case from some studies~\cite{DBLP:conf/kbse/GerasimouTC15,DBLP:journals/tse/WangHYY18,DBLP:journals/tsc/ChenB17}, where the aim is to optimize both reliability and latency of the SAS. We use one subject SAS only, as our goal here is to prove the existence of the likely issues when justification of indicators dose not aligned with the preferences in the SAS problem. Now, suppose that there are two solution sets \texttt{A} and \texttt{B} in Figure~\ref{fig:qi-example} returned by two search algorithms, and that 100\% reliability is of more interest to the engineer. To compare these two sets, a typical practice is to consider one or several commonly-used quality indicators, e.g., using those common ones in SBSE for SASs from Table~\ref{tb:taxonomy}. Since there is a strong preference towards 100\% reliability, the solution ($\beta$) of \texttt{B}, which reaches 100\% and has a lower cost than the corresponding one in \texttt{A}, should be the most ideal solution. However, the results from the quality indicators are quite opposite --- all the nine indicators evaluate \texttt{A} better than \texttt{B}, because all these quality indicators work on the assumption that the two objectives are incomparable and there is no preference of the problem, 
i.e., ignoring that the reliability needs to reach 100\% in this particular case. 
This is a typical example that indicates the risk of directly using existing quality indicators without considering the preferences of the problem~\cite{DBLP:conf/icse/Li0Y18} --- a major threat to $L_2$ level of justification.



\subsubsection{Suggestion and Opportunity}

The disappointment mentioned in Section~\ref{sec:rq3-di} can be mitigated by a simple suggestion:

\begin{tcolorbox}[breakable,left=5pt,right=5pt,top=5pt,bottom=5pt] 
\textbf{Suggestion 3:} Generic quality indicators should be adopted in conjunction with other evaluation methods for Pareto search on SASs. The justification needs to be made on the quality aspect that the indicator(s) cover and with respect to the preferences of the SAS problem considered (i.e., $L_1$). 
\end{tcolorbox}

To this end, a specific research opportunity raised is:

\begin{tcolorbox}[breakable,left=5pt,right=5pt,top=5pt,bottom=5pt] 
\textbf{Opportunity 3:} Preferences driven Pareto search for SASs. 
\end{tcolorbox}


Unlike weighted search where preferences need to be defined precisely to aim for a single optimal point in the objective space, the nature of Pareto search permits to take vague and imprecise preferences into account. For example, searching for a particular region in the objective space. This is of high interest in SAS where the nature of requirement specification is often imprecise~\cite{DBLP:journals/re/WhittleSBCB10}. More importantly, according to the SAS problems, engineers might only be interested in a handful of certain solutions that meet their preferences most in the objective space, instead of the entire set of trade-off solutions~\cite{DBLP:conf/icsa/CalinescuCGKP17,DBLP:journals/tosem/ChenLBY18}.

In this regard, preference information can be elicited and integrated as part of a generic quality indicator, or even serve as part of the fitness that drives the search towards the region of interest along a preferred direction. 
For example, preference information can be extracted from existing SAS design models or language, e.g., the Goal Model or RELAX~\cite{DBLP:journals/re/WhittleSBCB10} which contains a formal expression of preferences such as \textit{the latency shall be low while the cost shall ideally be as low as 5\$}. Next, it is possible to sample a vector of reference points, each dimension of which represents the expectation at the corresponding objective aligned with those preferences~\cite{DebSBC06,LiCMY18}. The resulted reference points could be directly exploited by a search algorithm or integrated into an indicator to assess the solution set thereafter.  

To achieve preferences driven Pareto search for SASs, there are several challenging research questions needed to be addressed:
\begin{itemize}

    \item How to (automatically) extract the preference information about the SAS problem in an efficient and cost-effective manner.
    

	\item How to structuralize the preferences of the SAS problem in a way that can be well reflected in the fitness of the search algorithm.
    
    \item What parts of the preferences, expressed in some software engineering representations, can be correlated to which quality aspect of the solution set.

\end{itemize}

\begin{figure}[!t]
  \centering
\begin{tikzpicture}
\begin{axis}[
  ybar=0.05cm,
  ymin=0,
  ymax=14,
    bar width=8pt,
  width=13cm,
    height=5cm,
    legend style={at={(0.4,0.96),font=\large},
      anchor=north,legend columns=2},
         y label style={at={(-0.08,0.5)},font=\large},
    ylabel={\# primary studies},
     nodes near coords,
     nodes near coords={\pgfmathfloatifflags{\pgfplotspointmeta}{0}{}{\pgfmathprintnumber{\pgfplotspointmeta}}},
    symbolic x coords={2009,2010,2011,2012,2013,2014,2015,2016,2017,2018,2019},
       x tick label style={rotate=45,anchor=east,font=\large},
    y tick label style={font=\large},
    xtick=data
    ]


\addplot[fill=blue!50,draw=black] coordinates {(2009,2) (2010,3) (2011,5) (2012,5) (2013,5) (2014,7) (2015,8) (2016,8) (2017,9) (2018,10) (2019,12)};

\addplot[fill=white,draw=black] coordinates {(2009,1) (2010,2) (2011,0) (2012,2) (2013,2) (2014,2) (2015,4) (2016,3) (2017,3) (2018,7) (2019,4)};

\legend{Problem nature, SE/SAS domain expertise}

\end{axis}

\end{tikzpicture}
   \caption{Popularity evolution on the domain information in SBSE for SASs (All studies use as least problem nature).}
 \label{fig:domain-count}
\end{figure}
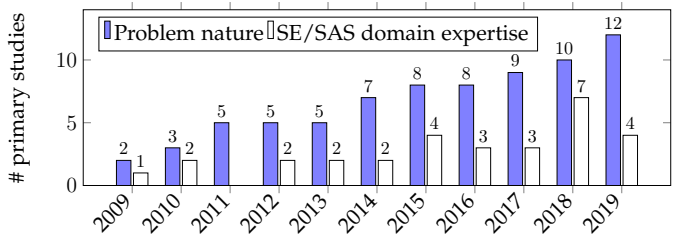

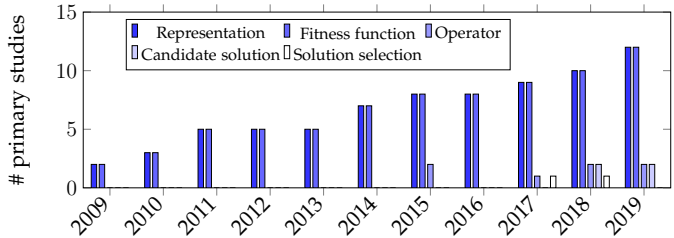
\begin{figure}[!t]
  \centering
\begin{tikzpicture}
\begin{axis}[
  ybar=0.05cm,
  ymin=0,
  ymax=15,
  width=13cm,
    height=5cm,
    bar width=3pt,
    enlargelimits=0,
     enlarge x limits=0.05,
    legend style={at={(0.4,0.96),font=\small},
      anchor=north,legend columns=3},
         y label style={at={(-0.08,0.5)},font=\large},
    ylabel={\# primary studies},
    symbolic x coords={2009,2010,2011,2012,2013,2014,2015,2016,2017,2018,2019},
       x tick label style={rotate=45,anchor=east,font=\large},
    y tick label style={font=\large},
    xtick=data
    ]

\addplot[fill=blue!80,draw=black] coordinates {(2009,2) (2010,3) (2011,5) (2012,5) (2013,5) (2014,7) (2015,8) (2016,8) (2017,9) (2018,10) (2019,12)};

\addplot[fill=blue!60,draw=black] coordinates {(2009,2) (2010,3) (2011,5) (2012,5) (2013,5) (2014,7) (2015,8) (2016,8) (2017,9) (2018,10) (2019,12)};

\addplot[fill=blue!40,draw=black] coordinates {(2009,0) (2010,0) (2011,0) (2012,0) (2013,0) (2014,0) (2015,2) (2016,0) (2017,1) (2018,2) (2019,2)};

\addplot[fill=blue!20,draw=black] coordinates {(2009,0) (2010,0) (2011,0) (2012,0) (2013,0) (2014,0) (2015,0) (2016,0) (2017,0) (2018,2) (2019,2)};

\addplot[fill=white,draw=black] coordinates {(2009,0) (2010,0) (2011,0) (2012,0) (2013,0) (2014,0) (2015,0) (2016,0) (2017,1) (2018,1) (2019,0)};

\legend{Representation, Fitness function, Operator, Candidate solution, Solution selection}

\end{axis}

\end{tikzpicture}
   \caption{Popularity evolution which parts of the search algorithms are specialized in SBSE for SASs (All studies specialize at least representation and fitness function).}
 \label{fig:spec-count}
\end{figure}

\subsection{RQ4: Specializing Search Algorithm for SASs}
\label{sec:rq4}

\subsubsection{Significance}

Most SBSE tasks inevitably require specializing the search algorithms in order to make them better serve the purpose, SAS problems are of no exception. It is, therefore, important to understand what, how, and why domain information of the SAS problems have been considered in such specialization when investigating SBSE for SASs. 

\subsubsection{Findings}

As mentioned, according to the categories proposed by Chen et al.~\cite{DBLP:journals/pieee/ChenBY20}, we first summarize the types of domain information used in SBSE for SASs. Figure~\ref{fig:domain-count} shows the results, anticipatedly, we see that problem nature is the fundamentally required domain information for the specialization in every study. This includes, for example, a variation point is categorical/numeric, the threshold constraint (e.g., full or partial satisfaction), and the scale/metric of the objective (e.g., worst or mean of the SAS's latency). SE/SAS domain expertise, in contrast, only forms the minority such as feature model~\cite{DBLP:journals/jss/PascualLPFE15,DBLP:journals/tosem/ChenLBY18}, adaptation tactics~\cite{DBLP:conf/icac/MorenoCGS16}, historical solutions~\cite{DBLP:conf/icse/KinneerCWGG18,DBLP:journals/infsof/ChenLY19}, and Markov model~\cite{DBLP:conf/icsa/CalinescuCGKP17}. Over the years, there is a tendency to widen the gap between the uptake of problem nature and SE/SAS domain expertise.

A more interesting question is perhaps which parts of a search algorithm have been specialized, regardless of whichever type of domain information used. From Figure~\ref{fig:spec-count}, we see that representation (e.g., a fixed-length vector or a tree) and fitness function (e.g., directly based on the SASs/simulator or derived from a well-defined mathematical model) are the essential parts in a search algorithm to be specialized. This is not surprising as they are always required in order to tailor a search algorithm to work on a SAS problem~\cite{DBLP:journals/csur/HarmanMZ12}. In contrast, there is only a small proportion of the studies (11 out of 74) that additionally consider other parts in the specialization, namely operators~\cite{DBLP:journals/tosem/ChenLBY18,DBLP:journals/jss/PascualLPFE15}, candidate solution~\cite{DBLP:conf/icse/KinneerCWGG18,DBLP:journals/infsof/ChenLY19} and solution selection~\cite{DBLP:conf/icsa/CalinescuCGKP17}, all of which involve SE/SAS domain expertise. For example, Kinneer et al.~\cite{DBLP:conf/icse/KinneerCWGG18} and Chen et al.~\cite{DBLP:journals/infsof/ChenLY19} leverage the good solution for a past problem instance or timestep to ``seed" the candidate solutions under the current search problem. The definition of goodness is entirely dependent on the SAS domain and engineering practices though. Calinescu et al.~\cite{DBLP:conf/icsa/CalinescuCGKP17} make use of Markov model for the SAS to define a boundary of robustness, which is then used to determine the survival of solutions during the search.

To provide details on how and why domain expertise is specialized, Table~\ref{tb:domain-reason} specifies all the types of SE/SAS domain expertise used, which parts of a search algorithm they have been specialized with and the reason behind. We can clearly see that the exploitation of the domain expertise all come with justifiable reasons, but their specializations may not go beyond the most fundamental representation and fitness function. Our findings for \textbf{RQ4} is therefore:

\begin{tcolorbox}[breakable,left=5pt,right=5pt,top=5pt,bottom=5pt] 
\textbf{Findings 10:} There is an increasing gap between the uptake of problem nature and SE/SAS domain expertise in SBSE for SASs \\
\textbf{Findings 11:} Regardless of whether problem nature or SE/SAS domain expertise has been used, nearly all the studies specialize in the representation and fitness function of a search algorithm only.
\end{tcolorbox}

\begin{table}[t!]
\centering
\caption{Reasons of leveraging SE/SAS domain expertise and their specializations in different parts of search algorithms on SBSE for SAS.}
\label{tb:domain-reason}

\centering

\begin{tabular}{cc}\toprule

\textbf{SE/SAS Domain Expertise}&\textbf{Reasons}\hfill\\ 
\midrule

Feature model $\Rightarrow$ R\&F (10)&\makecell{To systematically capture \\the variability of the software.}\\

\hline

Markov model $\Rightarrow$ R\&F (7)&\makecell{To provide formal verification\\ of the software states.}\\

\hline

Goal model $\Rightarrow$ R\&F (5)&\makecell{To better represent \\the stakeholders' needs.}\\

\hline
Tactics $\Rightarrow$ R\&F (5)&\makecell{To reduce the search space\\using prior expertise.}\\

\hline
\makecell{Abstract syntax tree\\$\Rightarrow$ R\&F (2)}&\makecell{To provide the most fundamental\\ sources that adapt the software.}\\

\hline
Feature model $\Rightarrow$ O (2)&\makecell{To comply with dependency\\and improve efficiency.}\\
\hline

Tactics $\Rightarrow$ O (2)&\makecell{To further restrict search\\space of adaptation.}\\
\hline
\makecell{Abstract syntax tree \\$\Rightarrow$ O (1)}&\makecell{To comply with code structure.}\\

\hline
Goal model $\Rightarrow$ O (1)&\makecell{To define what to search next\\ based on requirements.}\\
\hline
Seeding $\Rightarrow$ C (4)&\makecell{To improve convergence speed\\towards expected adaptation.}\\
\hline
Markov model $\Rightarrow$ S (2)&\makecell{To produce robust adaptation.}\\

\bottomrule
\end{tabular}

\begin{tablenotes}
    \footnotesize
    \item R, F, O, C and S denote representation, fitness function, operator, candidate solution and solution selection, respectively.
    \item Number in the bracket indicates how many studies are involved.
\end{tablenotes}
\end{table}

\subsubsection{Disappointments}
\label{sec:rq4-di}

Disappointingly, form the findings, we have failed to see how the advances of SBSE for SASs can be distinguished from ``yet another application domain of vanilla search algorithms", as SE/SAS domain expertise is often ignored (from Figure~\ref{fig:domain-count}) and the specialization in search algorithm rarely goes beyond the basic representation and fitness function under whichever type of domain information (from Figure~\ref{fig:spec-count})\footnote{Note that empirical studies, which may need to purposely compare the application of vanilla search algorithms for SAS, have been excluded.}. In other words, predominately the problem nature is used for the representation and fitness function of a search algorithm, representing a limited specialization.

Since SBSE is relatively new to SAS research, this result is predictable, but we did not expect such a significant gap and, as we have shown, the trend has no tendency to change over the years. Unlike the other disappointments discussed in this work, this disappointment may not cause immediate threats as the others but tend to have negative effects in the long-term. Indeed, a limited specialization may work without any issue for small and simple SASs, especially at the early dates. However, the SASs have now evolved to a stage with commonly high complexity, scales, dynamics and uncertainty (as we will show in Table~\ref{tb:sas}) that are hard to keep up with some of the assumptions made in vanilla search algorithms~\cite{DBLP:journals/jss/PascualLPFE15,DBLP:journals/tosem/ChenLBY18}. Therefore, ignoring the strong domain knowledge from engineers is a non-trivial issue in SBSE for SASs and can be an unwise waste of such valuable knowledge. At the same time, limiting specialization to only representation and fitness function could be harmful to the success of SBSE for SASs in the long-term~\cite{DBLP:journals/pieee/ChenBY20}. For example, searching without knowing the dependency relations in SAS may be difficult to find any valid solutions at all in the presence of complex dependencies~\cite{DBLP:journals/jss/PascualLPFE15,DBLP:journals/tosem/ChenLBY18}; producing only the non-dominated solutions while ignoring the robust ones is often undesirable in SAS design when there is an irregular trade-off surface~\cite{DBLP:conf/icsa/CalinescuCGKP17}. It is also not uncommon to see that similar concerns have been raised in the software engineering community, e.g., see Menzies's work~\cite{DBLP:journals/software/Menzies20}.

From our findings, we do see a few very good examples (e.g.,~\cite{DBLP:journals/tosem/ChenLBY18,DBLP:conf/icse/CailliauL17,DBLP:journals/infsof/ChenLY19,DBLP:journals/jss/PascualLPFE15}) on better specializing different parts of the search algorithms with SE/SAS domain expertise in SBSE for SASs. This is, in fact, a win-win strategy, where on the one hand, the search algorithm can be potentially made more controllable and explainable; on the other hand, the strong domain knowledge can serve as strong guidance to better steer different aspects of the search, achieving results that would be otherwise difficult to obtain. Further, the nature of complexity in SAS can actually provide more opportunity to design a better tailored and specialized search algorithm for the context. 

The ideal case would be specializing SE/SAS domain expertise in different parts of the search algorithms; if not, at least the SE/SAS domain expertise should be considered in the specialization or the problem nature should be exploited in parts other than representation and fitness function. This is what makes the work in SBSE for SASs rather unique and tailored to the SAS problems. In this way, we turn the search algorithms to be less general (i.e., typically not able to apply to other problems), but they are expected to work better (when being done properly) under the given SAS where the knowledge lies. Such an advanced specialization, albeit may not be essential, is often desirable in the long-term. Our disappointment is, therefore:

\begin{tcolorbox}[breakable,left=5pt,right=5pt,top=5pt,bottom=5pt] 
\textbf{Disappointment 4:} Limited specialization on search algorithms for SASs without tinkering with their internal designs.
\end{tcolorbox}

\begin{figure}[t!]
\centering
\includegraphics[width=\columnwidth]{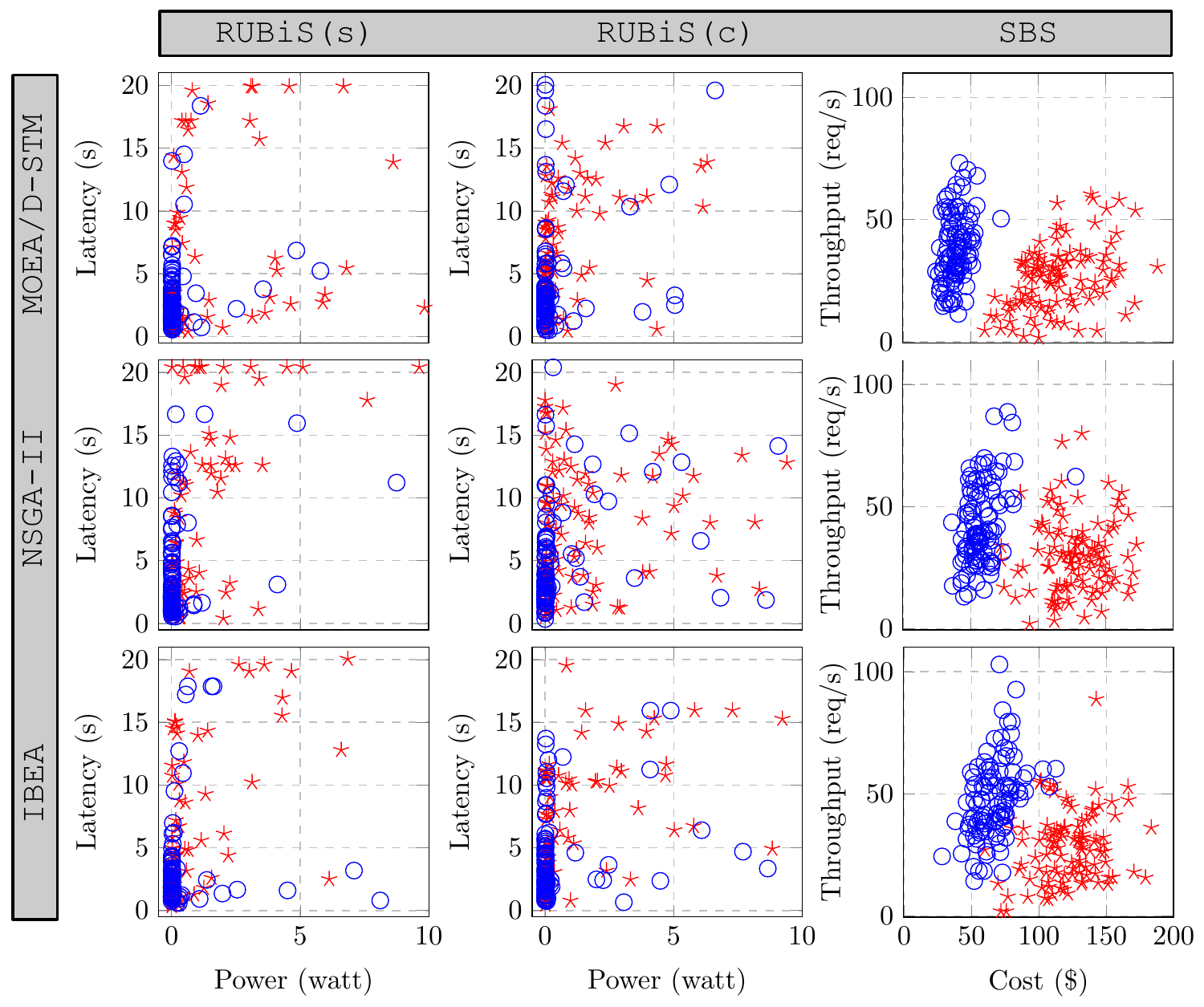}
\caption{Comparing the results between advanced (denoted as \copy\aMark) and limited (denoted as \copy\bMark) specialization on three search algorithms and three SASs under time-varying workload/services over 100 runs (Throughput is to be maximized while others are to be minimized).}
\label{fig:synergy}
\end{figure}

\subsubsection{Justification of the Likely Issues}

To justify the possible issue caused by limited specialization, we specialize Pareto search to optimize SASs that are designed using the feature model. The experiments contain three subject SAS, namely \textsc{RUBiS} with simple functionalities (\textsc{RUBiS(s)}), \textsc{RUBiS} with complex functionalities (\textsc{RUBiS(c)}), and a service-based system (\textsc{SBS}), all of which have been used in prior work~\cite{DBLP:conf/icac/MorenoCGS16,DBLP:conf/icac/GhahremaniG017,DBLP:journals/tosem/ChenLBY18}. The aim is to tune two objectives by adapting different variation points under time-varying workload and service quality/availability at runtime. Again, the reason for using three (two of which are from the same domain) is that, for SASs from the same or different domains, such a number is higher than 65\% and 73\% of the existing studies, respectively (as what will be shown in Section~\ref{sec:rq5}). We specialize in three search algorithms, i.e., MOEA/D-STM, NSGA-II, and IBEA, which are chosen because of their diverse characteristics and is the representative of their own kind. This is important for our justification, as we seek to showcase that the benefit of having better-specialized search can be generalized to different algorithms. We experiment 100 runs in total, under each of which a knee point is selected for self-adaptation. The settings of the algorithms and SASs details can be found in the supplementary. 

We compare two forms of specialization with identical search budget: (i) a limited one where no feature model is used, but the variation points and their types (e.g., numeric or categorical) are directly encoded as the representation. To enable efficient search at runtime, the fitness function is built by regression~\cite{DBLP:journals/tse/ChenB17} (for \textsc{RUBiS(s)} and \textsc{RUBiS(c)}) and the well-defined analytical model~\cite{DBLP:conf/gecco/0001LY18} (for \textsc{SBS}). As a result, it resembles a case where the vanilla search algorithm is used for the SASs. (ii) The advanced one that additionally parses the feature model, extracts only the critical features as the variation points in the representation and injects feature dependency into the reproduction operators of the search algorithm. In such a case, the SE/SAS domain expertise is the feature model while the specialization parts include the representation, fitness function, and operators. More details can be found from~\cite{DBLP:journals/tosem/ChenLBY18}.

From Figure~\ref{fig:synergy}, we see that the limited specialization, albeit does have some good results, often lead to solutions closer to the nadir points of both objectives or those that cause serve degradation on an objective with a little gain on the other. The advanced specialization, in contrast, performs overwhelmingly better than the limited counterpart, as its results have more points closer to the ideal region. The key rationale behind such a success is because the rich domain knowledge in the feature model is highly effective on which different parts of the search algorithm can rely.

\subsubsection{Suggestion and Opportunity}



Since the trend in SBSE for SASs has been on applying the vanilla version of the search algorithm(s) with only the compulsory amendments, our suggestion is, therefore:

\begin{tcolorbox}[breakable,left=5pt,right=5pt,top=5pt,bottom=5pt] 
\textbf{Suggestion 4:} Considering the possibility of tinkering with the vanilla search algorithm that is chosen justifiably for a SAS problem, especially the available SE/SAS domain expertise in relation to the internal algorithmic designs. 
\end{tcolorbox}

Indeed, the suggestion remains at a high level, but it can be centered as one thread of research opportunity:

\begin{tcolorbox}[breakable,left=5pt,right=5pt,top=5pt,bottom=5pt] 
\textbf{Opportunity 4:} Human-centric SBSE for SASs.
\end{tcolorbox}

\begin{table*}[t!]
\caption{Top 10 subject SAS to evaluate search algorithm and their characteristics on SBSE for SAS.}
\label{tb:sas}

\centering
\begin{tabular}{cccccccc}\toprule

\textbf{Subject SAS}&\textbf{Type}&\textbf{Domain}&\textbf{\# P}&\textbf{\# O.}&\textbf{Search Space}&\textbf{Env. change}&\textbf{Reasons}\\ 
\midrule

 Synthetic system&simulator&service&up to 15&1,2,3&up to $5.6\times 10^{18}$&concrete services&widely used (1), unknown (10)\\\hline
 RUBiS&real&web&10&1,2,4&up to $1.3\times 10^{16}$&workload&widely used (3), unknown (1)\\\hline
 UUV&simulator&vehicle&3-6&1,3,7&up to $3.9\times 10^{14}$&location&unknown (4)\\\hline
 Synthetic system&simulator&mobile&2,81&1,3&$2.4\times10^{24}$&workload&to assess GA (1), unknown (2)\\\hline
 Synthetic system&real&cloud&unknown&1,2&up to $10^{37}$&workload&unknown (3)\\\hline
 CrowdNav&simulator&navigation&7&1,2&$10^{65}$&traffic&unknown (3)\\\hline
 Travel system&real&web&78&8&$3.0\times 10^{23}$&workload&unknown (2)\\\hline
 Znn&real&web&unknown&1&unknown&workload&widely used (2)\\\hline
 WS-DREAM&data&service&up to 100&3&up to $3.4\times 10^{202}$&concrete services&widely used and realistic (2)\\\hline
 Online shopping&real&web&6&1,9&unknown&workload&unknown (2)\\

\bottomrule
\end{tabular}

\begin{tablenotes}
    \footnotesize
    \item Number in the bracket indicates how many studies are involved.
\end{tablenotes}
\end{table*}

The fundamental cause of the disappointments for \textbf{RQ4} is due to the lack of possible human involvement and their various forms of knowledge. This is, by design, often not part of a vanilla search algorithm. Traditionally, the main purpose of engineering SASs is to reduce the levels of human intervention on the running software systems. However, it has been shown that there are scenarios where human involvement is essential~\cite{DBLP:conf/icse/CamaraMG15}, or human knowledge has been proven to be able to greatly improve the behaviors of SASs. Similarly, SBSE is also motivated by the same origin: to automatically generate the software engineering process and thus free the software engineer from tedious and error-proven tasks. Recently, there is an ongoing demand to engineer human-centric SBSE~\cite{DBLP:journals/tse/RamirezRS19}, such that the search approach does not make the final decision on its own, but serving as an assistant to add insights for the human to make decisions. Those two facts, together, imply a perfect match between the two fields in terms of the human-centric aspect.

In particular, humans can refer to a wide range of engineers with certain software engineering expertise around SASs, including but not limited to, developers, requirements analysts, architects, and testers. Unlike classic SBSE for SASs, human-centric SBSE for SASs strikes for opening up the ``black box" of the vanilla search algorithm. A key outcome, when placing humans in the center of SBSE for SASs, is the tendency of encouraging more SE/SAS domain expertise uptakes, as the human in this context is all experts. In particular, it also promotes the specialization of different parts in the search algorithms, allowing humans to better explain and control the outcomes produced by a search algorithm.

A particularly interesting direction is the interactive SBSE for SASs, which enables the human to progressively learn and understand the characteristics of the SAS problem at hand and adjust towards more appropriate capture of the preferences information. As a result, the search can be driven to more precisely perform the expected behaviors, allowing the human to have more controllability over the search algorithm using their software engineering expertise on the SAS~\cite{LiCSY18,DebSKW10}. This would also create more inspirations to build specialized search algorithms, which should work the best under the SAS where the knowledge lies. The timely feedback retrieved from the search on SASs can also stimulate ``Innovization"~\cite{deb2014innovization} --- a particular situation where the SE/SAS domain expertise can also be consolidated as the search proceeds. Yet, some important challenges can be related to:

 \begin{itemize}

 \item What forms of SE/SAS knowledge/expertise can explicitly influence which aspects of SBSE for SASs.

 \item How humans can be placed in the loop of engineering SASs (either at design-time or runtime) in order to facilitate timely interaction with SBSE for SASs.

 \item How to ensure the information provided by humans is reliable, i.e., how to prevent immature inputs.

 \end{itemize}

\subsection{RQ5: Subject SASs in Evaluation}
\label{sec:rq5}

\subsubsection{Significance}

Research in SBSE for SASs would inevitably involve stochastic and random behaviors, either caused by the search algorithms and/or from the underlying SAS to be optimized. As a result, it is important to understand what, why, and how many subject SASs have been used in the evaluation.

\subsubsection{Findings}

Table~\ref{tb:sas} shows the top 20 subject SASs and their characteristics in SBSE for SASs. Clearly, we can see that they come from different types and domains, of diverse scales, the number of objectives, and search space, together with different environmental changes. An interesting finding is that a common reason behind the choice is because certain SASs are ``widely used"~\cite{DBLP:conf/icac/MorenoCGS16,DBLP:journals/tse/WangHYY18,DBLP:journals/taas/Garcia-GalanPTC16,DBLP:journals/jss/PascualLPFE15}. This is sensible as the purpose is often to generalize the findings on the most common SAS domains. A few of them, e.g., Pascual et al.~\cite{DBLP:conf/icse/PascualPF13}, have explicitly stated that the chosen SASs are particularly fit to evaluate the search algorithm studied (i.e., GA). However, there is still a considerable amount of the remaining studies that give no clear reasons for the choices.

\begin{figure}[!t]
  \centering
\begin{tikzpicture}
\begin{axis}[
  ybar=0.05cm,
  ymin=0,
  ymax=9,
    bar width=6pt,
  width=13cm,
    height=5cm,
    enlargelimits=0,
     enlarge x limits=0.05,
    legend style={at={(0.4,0.96),font=\large},
      anchor=north,legend columns=3},
         y label style={at={(-0.08,0.5)},font=\large},
    ylabel={\# primary studies},
     nodes near coords,
     nodes near coords={\pgfmathfloatifflags{\pgfplotspointmeta}{0}{}{\pgfmathprintnumber{\pgfplotspointmeta}}},
    symbolic x coords={2009,2010,2011,2012,2013,2014,2015,2016,2017,2018,2019},
       x tick label style={rotate=45,anchor=east,font=\large},
    y tick label style={font=\large},
    xtick=data
    ]

§

\addplot[fill=blue!50,draw=black] coordinates {(2009,2) (2010,1) (2011,3) (2012,1) (2013,3) (2014,4) (2015,4) (2016,3) (2017,4) (2018,7) (2019,8)};

\addplot[fill=white,draw=black] coordinates {(2009,0) (2010,2) (2011,2) (2012,2) (2013,1) (2014,2) (2015,3) (2016,4) (2017,5) (2018,4) (2019,3)};

\addplot[fill=red!50,draw=black] coordinates {(2009,0) (2010,0) (2011,0) (2012,2) (2013,1) (2014,1) (2015,2) (2016,1) (2017,1) (2018,0) (2019,1)};

\legend{Simulator,Real system,Data}

\end{axis}

\end{tikzpicture}
   \caption{Popularity evolution on the type of subject SAS to evaluate the search algorithms (Three studies use more than one type).}
 \label{fig:sas-count}
\end{figure}
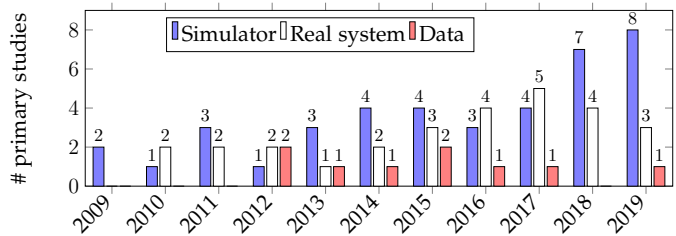

Another unique question in SBSE for SASs is what types of subject SAS are used. Overall, we found three types: real systems that involve actual deployment and running of the SAS; simulators that mimic the behaviors of a real system; and data that was collected from the real system but can be reconstructed and parsed to replicate the actual scenarios without the need to access the real system. Figure~\ref{fig:sas-count} shows the proportions of these three types with respect to the years. As can be seen, simulator is the most predominately used type and it exhibits an increasing popularity over the years, such as~\cite{DBLP:conf/icse/FredericksDC14,DBLP:conf/icse/Gerostathopoulos18,DBLP:journals/taas/ShevtsovWM19,DBLP:conf/saso/FredericksGK019}. Real system, in contrast, is much less commonly used, e.g.,~\cite{DBLP:journals/ase/GerasimouCT18,DBLP:journals/tosem/ChenLBY18,DBLP:journals/jss/PascualLPFE15}. Data is rarely adopted in the last decade~\cite{DBLP:journals/infsof/ChenLY19}. We found three studies~\cite{DBLP:journals/taas/LewisECRTY15,DBLP:conf/sigsoft/FilieriHM15,DBLP:conf/sigsoft/MaggioPFH17,DBLP:journals/tosem/ChenLBY18} where more than one type of SAS are used, and the reason for that is to improve the generalization of the results.

\begin{figure}[!t]
  \centering
  \begin{subfigure}[t]{0.5\columnwidth}
\begin{tikzpicture}[scale=0.8]
    \pie[
        /tikz/every pin/.style={align=center},
        text=pin,
        rotate=0,
        explode=0,
        color={white,black!65,black!55,black!45,black!35,black!25,black!15,black!5},
        ]
        {
            59/1,
            6/2,
            11/3,
            3/4,
            3/5,
            6/6,
          3/9,
           9/$\geq$10
        }
    \end{tikzpicture}
    \subcaption{Different settings or domains}
   \label{fig:subj-setting}
    \end{subfigure}
    \hspace{-0.2cm}
      \begin{subfigure}[t]{0.5\columnwidth}
\begin{tikzpicture}[scale=0.8]
    \pie[
        /tikz/every pin/.style={align=center},
        text=pin,
        rotate=0,
        explode=0,
        color={white,black!45,black!35,black!25,black!15,black!5},
        ]
        {
            73/1,
            14/2,
            6/3,
            3/4,
          1/9,
           3/12
        }
    \end{tikzpicture}
    \subcaption{Different domains only}
  \label{fig:subj-domain}
    \end{subfigure}
      \caption{Number of different subject SASs evaluated per study in SBSE for SASs.}
  \end{figure}
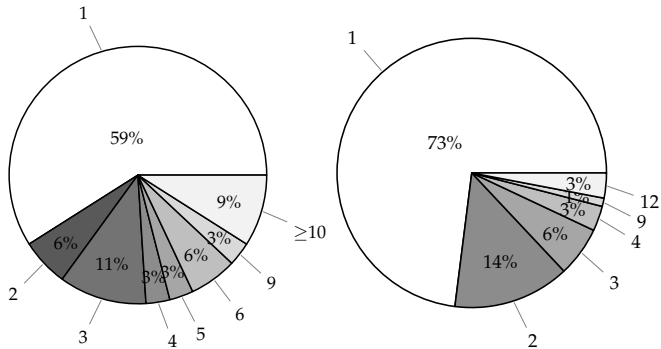

Indeed, it is important to evaluate the work on a set of subject SASs with different settings and/or from different domains~\cite{DBLP:conf/sigsoft/NagappanZB13,DBLP:conf/icse/SiegmundSA15}. To understand how many subject SASs are used per study, Figure~\ref{fig:subj-setting} shows the number of SAS with different settings or domains considered in a study. Interestingly, our result has indicated that there are 59\% primary studies consider only one SAS, a further 6\%, and 11\% consider two and three SASs, respectively.  Note that here, the SASs are differentiated based on settings, i.e., they are said different even if the study considers the same system under a given domain, as long as they have different structures, e.g., the same service-based systems with a different number of services to be composed from. If we differentiate the SASs solely based on their domains (e.g., a web system and an unmanned vehicle system), as shown in Figure~\ref{fig:subj-domain}, then the proportion of studies that consider one SAS increases to 73\%, and the number of studies that consider less than three subject SASs becomes 93\%. Our findings can be summarized as:

\begin{tcolorbox}[breakable,left=5pt,right=5pt,top=5pt,bottom=5pt] 
\textbf{Findings 12:} Various types of subject SASs have been used, with ``widely used" being the popular reason while often no justification is given. \\
\textbf{Findings 13:} Simulator is the most predominately used SAS type over years, followed by real systems.\\
\textbf{Findings 14:} Majority of the studies consider one subject SAS in the evaluation.
\end{tcolorbox}

\subsubsection{Disappointments}

Given the variety of subject SASs studied from the literature, it is disappointed to see that majority of the studies consider only one SAS in the evaluation. In particular, this is not because of a single overwhelmingly used benchmark, as shown in Table~\ref{tb:sas}. 

From the literature, the importance of diversity and coverage on subjects in evaluating software engineering research has been widely acknowledged. For example, sysmtematic studies conducted by Nagappan et al.~\cite{DBLP:conf/sigsoft/NagappanZB13} and Siegmund et al.~\cite{DBLP:conf/icse/SiegmundSA15} have concluded that:


\begin{displayquote}
``\emph{subject systems cover a wide range of different dimensions, which positively affects external validity.}"~\cite{DBLP:conf/icse/SiegmundSA15}
\end{displayquote}

As a result, a limited set of systems in the evaluation would inevitably weaken the generalization of conclusion, which is a major threat to external validity that remain unsolved in SBSE for SASs.

Admittedly, some studies, such as~\cite{DBLP:conf/icac/RamirezKCM09,DBLP:journals/jss/XuB19a,DBLP:conf/icse/KinneerCWGG18}, tend to provide an emerging idea together with a proof-of-concept evaluation only, in which case using one subject SAS might seem reasonable. We conjuncture that, however, it is not a sustainable trend for the research filed when such a proof-of-concept type of evaluation appears to be overwhelming, constituting the majority of the work from the literature. Our disappointment can then be summarized as:

\begin{tcolorbox}[breakable,left=5pt,right=5pt,top=5pt,bottom=5pt] 
\textbf{Disappointment 5:} Weak generalization of results across the subject SASs.
\end{tcolorbox}

\subsubsection{Justification of the Likely Issues}

To justify the likely issue on a limited number of subject SASs considered, we conduct experiments on both single-objective search (HC and RS) and Pareto search (MOEA/D and NSGA-II). These algorithms are chosen merely for illustration purposes and they are run with identical search budget, details the settings can be found in the supplementary. For Pareto search (for tuning latency, throughput, and cost), we run on the most widely used synthetic service-based systems, derived from the \textsc{WS-DREAM} dataset, with an aim to achieve runtime self-adaptation to a service change. We use up to four different workflows a in existing work~\cite{DBLP:conf/gecco/0001LY18,DBLP:journals/infsof/ChenLY19}, which offers better coverage than 76\% of the studies from the same domain as shown in Figure~\ref{fig:subj-setting}. The fitness is again evaluated by a well-defined analytical model~\cite{DBLP:conf/gecco/0001LY18,DBLP:journals/infsof/ChenLY19}. We use HV as the sole indicator because we seek to assess the overall quality of the solution set produced without specific preferences and it covers all quality aspects of a solution set. On the single-objective search, we chose six different SASs for design-time profiling scenarios. They are of diverse domains and have been used previously~\cite{DBLP:conf/icse/SiegmundKKABRS12,DBLP:journals/corr/abs-1801-02175}, which fits precisely with our goal of the justification. Note that according to Figure~\ref{fig:subj-domain}, six or more subject SASs from different domains has only been considered by 4\% of the studies. Details of theses SAS can be found in supplementary.

As shown in Table~\ref{tb:mo-rq5}, suppose that there are two sets of subject in the evaluation, \texttt{Set 1} with two subject SASs while \texttt{Set 2} with four. It is clear that \texttt{Set 1} could lead to a conclusion that NSGA-II being better. However, with a more thorough comparison in \texttt{Set 2}, we can understand that the conclusion from \texttt{Set 1} may not be the case: in fact, the two search algorithms gain competitive results. Similarly, for the single-objective case from Table~\ref{tb:so-rq5}, \texttt{Set 1} would imply that RS is better but \texttt{Set 2}, which involves a more extensive number of subject SASs, suggests that HC is actually better. The above exemplifies how a limited number of subject SAS can mislead the conclusion and weaken the generalization.

 \begin{table}[t!]
 
\centering
	\caption{Two subject sets to compare the 30 runs' mean HV (three objectives: latency, throughput and cost) by NSGA-II and MOEA/D on different SASs upon a service quality/availability change.}
\label{tb:mo-rq5}

\centering
\begin{center}
\begin{tabular}{ccc|cc}\toprule

\multirow{2}{*}{\textbf{SASs}}&\multicolumn{2}{c|}{\textbf{Subject Set 1}}&\multicolumn{2}{c}{\textbf{Subject Set 2}}\\
\cmidrule{2-5}

&\textbf{NSGA-II}&\textbf{MOEA/D}&\textbf{NSGA-II}&\textbf{MOEA/D}\\ 
\midrule

\textsc{WS-DREAM-1}&\cellcolor{yellow!50}0.9776&0.9413&\cellcolor{yellow!50}0.9776&0.9413\\

\textsc{WS-DREAM-2}&&&0.9662&\cellcolor{yellow!50}0.9758\\

\textsc{WS-DREAM-3}&\cellcolor{yellow!50}0.9705&0.9531&\cellcolor{yellow!50}0.9705&0.9531\\

\textsc{WS-DREAM-4}&&&0.9372&\cellcolor{yellow!50}0.9769\\



		







\bottomrule
\end{tabular}
\end{center}
\begin{tablenotes}
    \footnotesize
    \item The better one is highlighted. All comparisons are statistically significant ($p<$.05 on Wilcoxon signed-rank test) and with large effect sizes (on $A_{12}$).
\end{tablenotes}
\end{table}
 \begin{table}[t!]
 
\centering
	\caption{Two subject sets to compare the 100 runs' mean latency (s) optimized by Hill Climbing (HC) and Random Search (RS) on different domains of SASs under a workload change.}
\label{tb:so-rq5}

\centering
\begin{center}
\begin{tabular}{ccc|cc}\toprule

\multirow{2}{*}{\textbf{SASs}}&\multicolumn{2}{c|}{\textbf{Subject Set 1}}&\multicolumn{2}{c}{\textbf{Subject Set 2}}\\
\cmidrule{2-5}

&\textbf{HC}&\textbf{RS}&\textbf{HC}&\textbf{RS}\\ 
\midrule

\textsc{Apache}&&&\cellcolor{yellow!50}0.86&0.87\\

\textsc{BDBC}&0.41&\cellcolor{yellow!50}0.40&0.41&\cellcolor{yellow!50}0.40\\

\textsc{BDBJ}&&&\cellcolor{yellow!50}5.50&6.64\\

\textsc{LLVM}&203.88&\cellcolor{yellow!50}202.30&203.88&\cellcolor{yellow!50}202.30\\

\textsc{x264}&&&\cellcolor{yellow!50}244.33&247.82\\

\textsc{SQLite}&&&\cellcolor{yellow!50}14.99&15.52\\

\bottomrule
\end{tabular}
\end{center}
\begin{tablenotes}
    \footnotesize
    \item The better one is highlighted. All comparisons, except \textsc{BDBC}, are statistically significant ($p<$.05 on Wilcoxon signed-rank test) and with large effect sizes (on $A_{12}$).
\end{tablenotes}
\end{table}

\subsubsection{Suggestion and Opportunity}

Our suggestion is straightforward:

\begin{tcolorbox}[breakable,left=5pt,right=5pt,top=5pt,bottom=5pt] 
\textbf{Suggestion 5:} Aiming to evaluate SBSE for SASs work with at least two subject SASs, which can be with different settings (better coverage than 59\% of the studies) or from different domains (better coverage than 73\% of the studies). Ideally, the more subject SASs the stronger conclusion, but using only one subject should not be recommended.
\end{tcolorbox}

Indeed, it is easy to argue that \textit{``we need a higher number of subject SASs"}, but practically this depends on many factors, such as the resources to deploy, test, and configure a real SAS, as well as the time for the experiments to run. We do not attempt to undermine such an effort, as this is one of the difficulties that distinguishes research on SAS and many other fields. With this in mind, applying simulators can be an option\footnote{For example, the artifacts collection from SEAMS: \url{https://www.hpi.uni-potsdam.de/giese/public/selfadapt/exemplars/}.}, as relatively they are often simpler to be deployed and can run faster. However, simulators often rely on a fixed assumption about the realistic scenarios, which may not hold. Therefore, conducting research in SBSE for SASs using the simulator could pose threats to construct validity. 

Indeed, a possible solution could be to use real SASs as the complement to the simulator. This, however, does not solve our disappointment: it merely separates the evaluation on two hierarchies, where on the real SAS part the number of subject SAS may still be small due to the cost of setting up the experiments. As a result, the conclusions drawn would be biased towards the simulation part. The key is how to retain sufficient realism while keeping the efforts low. From this perspective, the realistic data, which may be easily parsed while is collected from real SAS, can be a promising source. However, as our results form Figure~\ref{fig:sas-count} indicated that, over years, there is not much readily available dataset for SAS. An opportunity for \textbf{RQ5} is, therefore:

\begin{tcolorbox}[breakable,left=5pt,right=5pt,top=5pt,bottom=5pt] 
\textbf{Opportunity 5:} Reusable real-world dataset collection and sharing in SBSE for SASs.
\end{tcolorbox}

A unique property of the dataset for evaluating SBSE on SAS is that it needs to involve certain complexity of the search, e.g., search space, number of objectives, or number of variation points. The collection process of the data itself, as expected, would be expensive. However, once such data has been collected, it can benefit the community as a whole for possible reuse and benchmark. We, therefore, call for the community to join the effort on building an ecosystem of collecting and maintaining real-world data for SAS, based on which perhaps more realistic simulators can be built. Some specific challenges are:

 \begin{itemize}

 \item How to define environmental conditions for a SAS during the data collection process?

 \item How to sample the variation point and the objective?

 \item How to codify a data collection protocol that can mitigate measurement bias?

 \end{itemize}

\section{Other Opportunities}
\label{sec:opp}

Apart from the opportunities discussed under each of the RQs, we have also unidentified other opportunities which are promising to promote SBSE for SASs but are unfortunately under-explored. In what follows, we elaborate on these opportunities in detail.

\subsection{Effective and Efficient Fitness Evaluation in SBSE for SASs}
A crucial part of SBSE is how the fitness of a solution can be evaluated, which serves as the key to driving the search process. This, in the context of SASs, is often related to how the behaviors of the systems can be changed with different adaptation solutions. In certain scenario, it is possible to profile the SAS at design-time, or at runtime where the profiling only affects the SAS in certain aspects rather than changing the whole system~\cite{DBLP:conf/icse/Gerostathopoulos18}. However, most commonly, such profiling is expensive and time-consuming. In contrast, surrogate models that are based on machine learning has been explored as an alternative, given that they are relatively cheap in terms of the fitness evaluation as the search proceeds~\cite{DBLP:conf/icse/Chen19b}. Yet, this comes with the cost of high complexity in building such a model, which may still be lack of accuracy or difficult to capture the up-to-date changes of SASs. Further, the number of examples required to train the model can also hinder the effectiveness of the search.

The situation raises the research opportunity of investigating effective and efficient fitness evaluation in SBSE for SASs. In particular, the key difficulty lies in the question of how to keep the overhead of fitness evaluation low, while maintaining a reasonable accuracy and cost of building the surrogate model. A promising direction on this is the research area of incremental online learning, where the model can be learned with limited data samples and can be efficiently updated as new data is collected while providing adequate accuracy~\cite{DBLP:conf/icse/Chen19b}. The other possible direction is to explore the so-called novelty model that does not require to observe the behaviors of SASs when using SBSE~\cite{DBLP:conf/kbse/RamirezJCK11}. Such a model mimics the natural phenomenon where the evolution would never be solely guided by explicit objectives, but also the biological novelty of the individuals. In such a way, the fineness can be assessed without the need to affect or acquire data from the SASs, and thus mitigating expensive evaluation. However, more research questions need to be addressed in order to better incorporate online learning with SBSE for SASs, such as the following: 

\begin{itemize}

\item Whether the frequency of model updates could have an impact on the search results.

\item How to handle the trade-off between the cost of model building and the accuracy (or relevance) of the model, if any.

\item What are the correlations between the accuracy (or relevance) of a model to the improvement of SBSE for SASs.

\end{itemize}

\subsection{Just-in-Time Handling of Changes in SBSE for SASs}

SAS would inevitably face changes in the requirements, environment, or its internal states, either at design-time or at runtime. Despite the fact that SBSE is capable of naturally handling dynamics to some extent, the more fundamental problem is how often should the optimization runs in order to ensure that the results can cope with the new changes. Current researches on SBSE for SASs have almost ignored this point or simply assumes that the search algorithm can be re-triggered when there is a need (e.g., according to a fixed frequency or upon the occurrence of changes). Yet, such a strategy would suffer the limitation that no changes can be captured during the run of the search algorithm.

To this end, recent advances on so-called dynamic optimization~\cite{DBLP:journals/swevo/NguyenYB12} and dynamic SBSE~\cite{harman2012dynamic} is a promising but under-explored solution for SASs. Here, the key idea is to allow the search algorithm to automatically pick up any new changes during the search process, and therefore the new information can be used to steer the search or old and useless information can be discarded in order to prevent misleading. Such a very nature is a perfect fit for various problems with ``changes" that are faced by modern SASs. However, there are some crucial challenges in this particular direction of research on SBSE for SASs, for example:

 \begin{itemize}

 \item What are the mappings between the changes in SASs and the changes with respect to the search algorithm.

 \item What are the changes can be handled while the search is under processing, and how they can be fed into the search.

 \item Whether it is possible to generically consolidate any given search algorithm.

 \end{itemize}

 \subsection{Incorporating SBSE with Other Approaches for SASs}

SBSE would never be the sole approach for tackling problems in SASs. In fact, given the nature of ``optimization" implied in SBSE, there is a variety of opportunities to incorporate SBSE and other approaches for SASs, such as control theory, verification, machine learning, and so forth. Our review has witnessed a few successful works that specifically incorporate SBSE with the other approaches. For example, Maggio et al.~\cite{DBLP:conf/sigsoft/MaggioPFH17} have applied control-theoretic adaptation whose internal control signals are optimized by using SBSE. In general, however, there is a lack of generic guidelines about the possible forms of incorporation. This is important, especially given the wide applicability of SBSE and other approaches for engineering SASs. In particular, challenges can be raised by the following new directions of research:


 \begin{itemize}

 \item What are the patterns involved when incorporating SBSE with the other approaches for engineering SASs.

 \item Whether there could be a ``symbiotic" relation exist between SBSE and another approach, i.e., both SBSE and the other can benefit from each other, which collaborates together to improve the SAS.

 \item How to codify a generic methodology that guides the practitioners of SASs on incorporating SBSE with the other approaches.

 \end{itemize}

\section{Threats to Validity}
\label{sec:tov}

Threats to construct validity can be raised by the research methodology, which may not serve the purpose of answering our research questions. We have mitigated such threats by following the systematic review protocol proposed by Kitchenham et al.~\cite{DBLP:journals/infsof/KitchenhamBBTBL09}, which is a widely recognized search methodology for conducting a survey on software engineering research. Another threat is related to the citation count used in the exclusion criteria. Indeed, 
    it is difficult to set a threshold for such, 
    as the citation count itself cannot well reflect the impact of work, thereby such exclusion criteria can be harmful to the construct validity.
    It is however worth noting that our goal is to analyze the major trends about how SBSE has been used for SASs, which can at least provide some sources for analyzing and justification. Further, it is necessary to reach a trade-off between the trend coverage and the efforts required for detailed data collections of the studies. Of course, the citation from Google Scholar could be biased by its underlying mechanism, but it remains uncertain about which online repository offers the most reliable citation information.
    

Threats to internal validity may be introduced by having inappropriate classification and interpretation of the papers. We have limited this by conducting multiple rounds of paper reviews amongst all the authors. Error checks and investigations were also conducted to correct any issues found during the search procedure. Another related threat to internal validity is that there was a considerable gap between the completion of collection and the submission/final publication, and therefore it raises a timeliness issue, particularly with respect to the citation count used in the exclusion criteria. This is, however, not uncommon for all survey studies and hence remains an open problem. Another threat is caused by information that has not been stated in the studies. For example, a possible reason for using a search algorithm could be that it is the only one with readily available implementation, but none of the studies has stated this clearly.

Finally, threats to external validity may restrict the generalization of the results. We have mitigated such by conducting the systematic survey wider and deeper: it covers 3,740 searched papers published between 2009 and 2019, on 27 venues from 7 repositories; while at the same time, extracting 74 most notable primary studies following the exclusion and inclusion procedure.

\section{Conclusion}
\label{sec:con}
In this work, we have systematically surveyed the research on SBSE for SASs published between 2009 and 2019, leading to a large set of studies span across 27 venues, based on which 409 ones were identified for detailed review and eventually 74 primary studies were selected for the analysis. Several key statistics have been extracted from the state-of-the-art with respect to the RQs:

\begin{itemize}
\item \textbf{To RQ1:} In the past decade, LS, GA, and IP solver are the most popular search algorithm on the single/aggregated objective case. NSGA-II is predominant for Pareto search. Their justification of choice are mainly at $L_3$ or $L_4$, despite they are used in a different context of SASs.
\item \textbf{To RQ2:} Single objectives are less commonly assumed than its multiple objective counterparts, within which weighted search is predominant over the years. The actual objectives to be searched are varied, but latency and cost are of the widest concern.
\item \textbf{To RQ3:} On Pareto search, the raw objectives are most commonly used in the evaluation and a considerable amount of studies have used no generic quality indicator at all, without justification. For those that do use, the justification of choices is mainly at level $L_2$ or $L_3$. This is a consistent trend across the years.
\item \textbf{To RQ4:} There is an increasing gap between the uptake of problem nature and SE/SAS domain expertise, while most studies specialize in the representation and fitness function of a search algorithm only.
\item \textbf{To RQ5:} Over the years, simulators are the most commonly used types of subject SASs and the majority of the studies consider only one subject SAS, regardless of the settings and domains.
\end{itemize}

The results have also revealed five disappointments from the most notable primary studies, namely:

\begin{itemize}
\item Unjustified bias on the choice of search algorithms.
\item Unjustified and limited formulation on the multi-objective search for SASs.
\item Questionable choice of evaluation methods in Pareto search for SASs.
\item Limited specialization on search algorithms for SASs without tinkering with their internal designs.
\item Weak generalization of results across the subject SASs.
\end{itemize}

We present theoretical and/or experimental evidence to justify the issues, provide suggestions, and also highlight eight emergent opportunities that are currently under-explored for research on SBSE for SASs, theses are:

\begin{itemize}
\item Generic guidance on justifiably choosing search algorithm(s) according to the requirements of the particular SAS problem studied.
\item Pareto many-objective search for SASs.
\item Preferences driven Pareto search for SASs.
\item Human-centric SBSE for SASs.
\item Reusable real-world dataset collection and sharing in SBSE for SASs.
\item Effective and efficient fitness evaluation in SBSE for SASs.
\item Just-in-time handling of changes in SBSE for SASs.
\item Incorporating SBSE with other approaches for SASs.
\end{itemize}

Our work provides useful insights that can hopefully excite a much more significant growth of this particular field of research, attracting not only the SAS practitioners but also the researchers from the other fields, such as general SBSE, Computational Optimization, and Evolutionary Computation.

\ifCLASSOPTIONcaptionsoff
  \newpage
\fi



%

\section*{Acknowledgement}
We thank the doctoral researchers from the IDEAS laboratory at Loughborough University for their assistance in collecting and analyzing the data in this work. We also would like to thank all the anonymous reviewers for their constructive comments that help to significantly improve this paper.

\bibliographystyle{IEEEtran}
\bibliography{ref}

\end{document}